\documentclass[aps,pre,twocolumn,showpacs,groupedaddress,superscriptaddress, amsmath,amssymb]{revtex4-1}

\usepackage[pdftex]{graphicx}

\usepackage{amsmath}
\usepackage{amssymb}
\usepackage{amsfonts}
\usepackage{dcolumn}
\usepackage{bm}
\usepackage{color}

\begin{document}

\title{Density of states and dynamical crossover in a dense fluid revealed by exponential mode analysis of the velocity autocorrelation function}

\author{S. Bellissima}
\affiliation{Dipartimento di Fisica e Astronomia, Universit\`a degli Studi di Firenze, via G. Sansone 1, I-50019 Sesto Fiorentino, Italy}

\author{M. Neumann}
\affiliation{Fakult\"{a}t f\"{u}r Physik der Universit\"{a}t Wien, Strudlhofgasse 4, A-1090 Wien, Austria}

\author{E. Guarini}
\affiliation{Dipartimento di Fisica e Astronomia, Universit\`a degli Studi di Firenze, via G. Sansone 1, I-50019 Sesto Fiorentino, Italy}

\author{U. Bafile}
\thanks{Corresponding Author. Tel.: +39-055-522-6676; Fax: +39-055-522-6683}
\email{ubaldo.bafile@isc.cnr.it}
\affiliation{Consiglio Nazionale delle Ricerche, Istituto dei Sistemi Complessi, via Madonna del Piano 10, I-50019 Sesto Fiorentino, Italy}

\author{F. Barocchi}
\affiliation{Dipartimento di Fisica e Astronomia, Universit\`a degli Studi di Firenze, via G. Sansone 1, I-50019 Sesto Fiorentino, Italy}

\begin{abstract}

Extending a preceding study of the velocity autocorrelation function (VAF) in a simulated Lennard-Jones fluid [Phys. Rev. E {\bf 92}, 042166 (2015)] to cover higher-density and lower-temperature states, we show that the recently demonstrated multi-exponential expansion method allows for a full account and understanding of the basic dynamical processes encompassed by a fundamental quantity as the VAF. In particular, besides obtaining evidence of a persisting long-time tail, we assign specific and unambiguous physical meanings to groups of exponential modes related to the longitudinal and transverse collective dynamics, respectively. We have made this possible by consistently introducing the interpretation of the VAF frequency spectrum as a global density of states in fluids, generalizing a solid-state concept, and by giving to specific spectral components, obtained through the VAF exponential expansion, the corresponding meaning of partial densities of states relative to specific dynamical processes. The clear identification of a high-frequency oscillation of the VAF with the near-top excitation frequency in the dispersion curve of acoustic waves is a neat example of the power of the method. As for the transverse mode contribution, its analysis turns out to be particularly important, because the multi-exponential expansion reveals a transition marking the onset of propagating excitations when the density is increased beyond a threshold value. While this finding agrees with the recent literature debating the issue of dynamical crossover boundaries, such as the one identified with the Frenkel line, we can add detailed information on the modes involved in this specific process in the domains of both time and frequency. This will help obtain a still missing full account of transverse dynamics, in both its nonpropagating and propagating aspects which are linked through dynamical transitions depending on both the thermodynamic states and the excitation wavevectors. 

\end{abstract}


\date{\today}

\maketitle

\section{Introduction}
\label{sect: intro}

The velocity autocorrelation function (VAF) is a key quantity in the translational dynamics of a fluid at the microscopic scale \cite{Hansen, Balucani, Boon}. Phenomena such as thermal and mass diffusion, sound propagation, transverse-wave excitation, having either a single-particle or a collective nature, are all reflected, through the motions of individual particles, in a fundamental quantity, which, in the usual statistical-mechanical treatment of a many-body system, is expressed in terms of the time autocorrelation function of an atomic variable, namely the particle velocity \cite{Gaskell1, Gaskell2}.

On its own, the VAF is not a directly measurable quantity, and its investigation has been for a long time a topical subject of theoretical and simulation studies \cite{Balucani, Boon, Levesqueverlet}. It is well known that a major reason for studying the VAF was the detection, in a molecular dynamics (MD) simulation of a hard-sphere (HS) fluid at intermediate density, of a long-time tail (LTT) well accounted for by a power law $A\,t^{-3/2}$ \cite{Alder_vaf, Alder_LTT}.

A number of simulation studies were later devoted to fluids with more realistic interaction potentials, in order to establish the presence and the time behavior of the LTT depending on the thermodynamic conditions of the fluid \cite{Levesque, McDonough, Dib, Meier1, Meier2, Williams, Ryltsev}. It was found that the LTT was most easily detected at intermediate densities in the gaseous state, while in dense liquids other dynamical effects on shorter time scales, such as backscattering due to the bouncing of atoms off near neighbours, effectively hide the LTT \cite{McDonough}.

On the theoretical side, both hydrodynamics-based arguments \cite{Kawasaki, Ernst1970, Ernst1971} and kinetic theory for HS fluids \cite{Dorfman} led to the time dependence $A\,t^{-3/2}$ and provided the same expression for the amplitude coefficient $A$ [see Eq. (\ref{eq:A-balucani}) below]. However, as recently recalled \cite{paper1}, the two treatments lead to different results for the asymptotic VAF time dependence. Moreover, the theoretical predictions have not been confirmed by a two-dimensional simulation of a very large system of hard disks \cite{Isobe}.

Although so much attention was paid to the LTT issue, it is clear, however, that the VAF contains relevant information about the atomic dynamics at all time scales, starting from the very-short-time binary collision processes. Therefore, a complete approach to the VAF analysis should cover in a consistent way the full time range from zero up to the LTT regime in any thermodynamic state of the fluid.

In a recent paper \cite{paper1}, this goal has been attained, for a Lennard-Jones (LJ) fluid at a supercritical temperature and intermediate densities, by exploiting a general property of correlation functions of dynamical variables in a many-particle Hamiltonian system, whose time dependence has been recently shown to be represented as infinite series of (complex and/or real) exponentials \cite{Barocchi_2012, Barocchi_2013, Barocchi_2014}. In summary, it was found \cite{paper1} that, for all the considered thermodynamical states considered and in the whole time range accessed by MD, the VAF is described perfectly by fitting the sum of a small number of exponential terms. Four of them are real exponentials having decay times ranging from the order of one collision time (``fast modes'') up to the order of several tens of collision times (``slow modes''). The slowest-decay exponential, and to a lesser extent the second-slowest one, together build up a long-time behavior which reproduces the LTT at least as accurately as the power-law dependence. In addition to the real ones, other terms present in the fit function are two pairs of complex conjugate exponentials representing damped oscillations. Each of them is characterized by a decay time, also of the order of, or less than, one collision time, and a frequency close to the collision frequency. These modes represent fast oscillatory components of the VAF and correspond to vibratory motions of the particles.

All the parameters describing the various modes represented by exponential terms display a completely smooth density dependence, signaling a continuous evolution of all aspects of the atomic dynamics. Nevertheless, one observes that the intensity of the complex modes grows with density at the expense of the fast real modes. This was then naturally interpreted as the evolution towards the formation of better structured nearest-neighbor cages favoring oscillatory motions, though strongly damped. It has to be noted, however, that this picture still refers to the case where the VAF displays a behavior typical of a medium-density gas rather than that of a liquid, and shows, for example, a monotonic decrease without reaching negative values.

The effectiveness of this analysis in allowing for a simple and consistent description of the entire time dependence of the VAF has naturally suggested to extend this approach to thermodynamic states closer to a liquidlike, and even true-liquid, behavior, for a number of reasons. One is the fact that it is still unclear whether the LTT persists in denser and/or colder fluids, given that the overall decay to zero reduces the VAF intensity to very small values in the time range where the LTT should be its dominant feature.

However, a more important point is the above mentioned change of the VAF shape and sign. So far, no theory has produced a model able to describe, at the same time and in an accurate way, both the negative portion of the VAF and its subsequent growth to positive values, followed by the final decay to zero from above. This is the time range where oscillatory components of the total VAF develop when the dynamical behavior tends to the one typical of a liquid.

The interpretation of the frequency spectrum of the VAF as a density distribution of states, which is straightforward for a crystalline solid, when extended to the fluid phase provides, in principle \cite{Gaskell1, Gaskell2}, a route to the investigation of the dynamics also for what concerns the collective motions. However, no clear way of carrying out this task has been developed so far. 

It is, thus, a main goal of this work to explore the possibility of exploiting the mode decomposition provided by the exponential expansion in order to relate the various VAF components to specific collective and single-particle motions typical of a fluid.

We have, therefore, performed more MD simulations in order to extend the analysis carried out in Ref.\ [\onlinecite{paper1}] both to the high density supercritical fluid and to some liquid states at subcritical temperatures. The results of this analysis will show that the evolution of the fitted exponential terms in number, nature, intensity and time scale, gives indeed valuable insight on the identification of some of the fitted modes with specific dynamical properties. In particular, besides the already recognized role of slowly decaying exponentials in accounting for the LTT \cite{paper1}, which is confirmed by the new simulations, we also demonstrate the correspondence between the highest-frequency complex mode and the dispersion curve of longitudinal acoustic wave propagation. Moreover, we will show that the remaining exponential modes represent the transverse collective dynamics and that their density evolution provides clear evidence of a crossover.

In this context it is worth to mention recent studies \cite{Simeoni, Gorelli, Bolmatov, Cunsolo_2016} that have discussed the existence of a transition between a low-density gaslike and a high-density liquidlike dynamical regime of a supercritical fluid. This crossover was identified by Brazhkin and coworkers \cite{Brazhkin_2012} with the crossing of the so-called Frenkel line, which has been eventually connected directly to the onset of a nonmonotonic time dependence of the VAF \cite{Brazhkin_2013}.
 
In Sect.\ \ref{sect: simul} we present the results of the new simulations. After a brief summary of the exponential mode analysis (Sect.\ \ref{sect: multiexp}), its results are reported and discussed in Sect.\ \ref{sect: results}. The conclusions are summarized in Sect.\ \ref{sect: concl}.  
 
\section{Simulations}
\label{sect: simul}

\begin{figure}
\resizebox{0.65\textwidth}{!}
{\includegraphics[viewport=6cm 18.5cm 17.5cm 25.5cm]{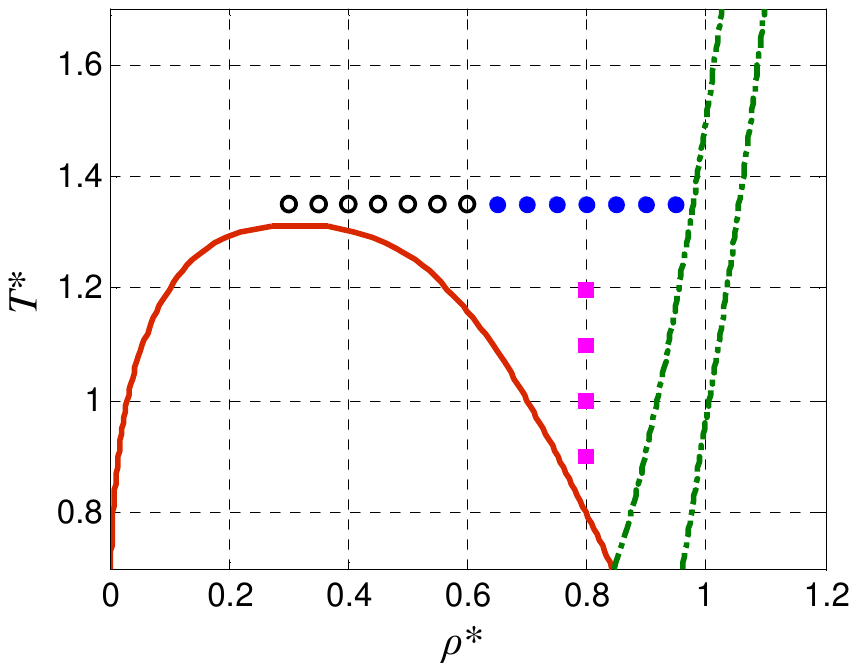}}
\caption{(Color online) Simulated LJ states in the $(\rho^*,T^*)$ plane. Black open circles indicate the points investigated in Ref. [\onlinecite{paper1}]. The other points refer to this work: blue dots are higher density states on the same $T^*=1.35$ isotherm, while pink squares are at lower temperatures along the $\rho^*=0.80$ isochore. Also shown are the liquid-vapor coexistence line (red, full line) \cite{Johnson}, and the two boundaries of the solid-fluid coexistence region (green, dot-dashes) \cite{Mastny}.}
\label{fig:thermodyn} 
\end{figure} 

MD simulations of the VAF for the LJ (12--6) fluid along the slightly supercritical $T^*=1.35$ isotherm were already reported for particle number densities up to $\rho^*=0.60$ \cite{paper1}. Here we consider higher density states (with $\rho^*$ varying from 0.65 to 0.95 in steps of 0.05) at the same temperature, reaching a density very close to that where isothermal solidification begins. Moreover, we simulated a few colder liquid states along the $\rho^*=0.80$ isochore and below the critical isotherm, with $T^*=1.20, 1.10, 1.00, 0.90$. The point at $T^*=1.35$, $\rho^*=0.80$ thus belongs to both the isotherm and the isochore here considered. Throughout this paper, asterisks denote standard reduced variables, like $T^*=k_{\rm{B}}T/\epsilon$ and $\rho^*=\rho\sigma^3$, where $k_{\rm B}$ is the Boltzmann constant and $\epsilon$ and $\sigma$ are the energy and length scale parameters of the LJ potential. The particle mass will be denoted by $m$. Figure \ref{fig:thermodyn} shows the state points in the $(\rho,T)$ plane, including for completeness and reference also the lower density states of Ref.\ [\onlinecite{paper1}]. 

\begin{figure}
\resizebox{0.65\textwidth}{!}
{\includegraphics[viewport=6.5cm 16cm 16cm 25cm]{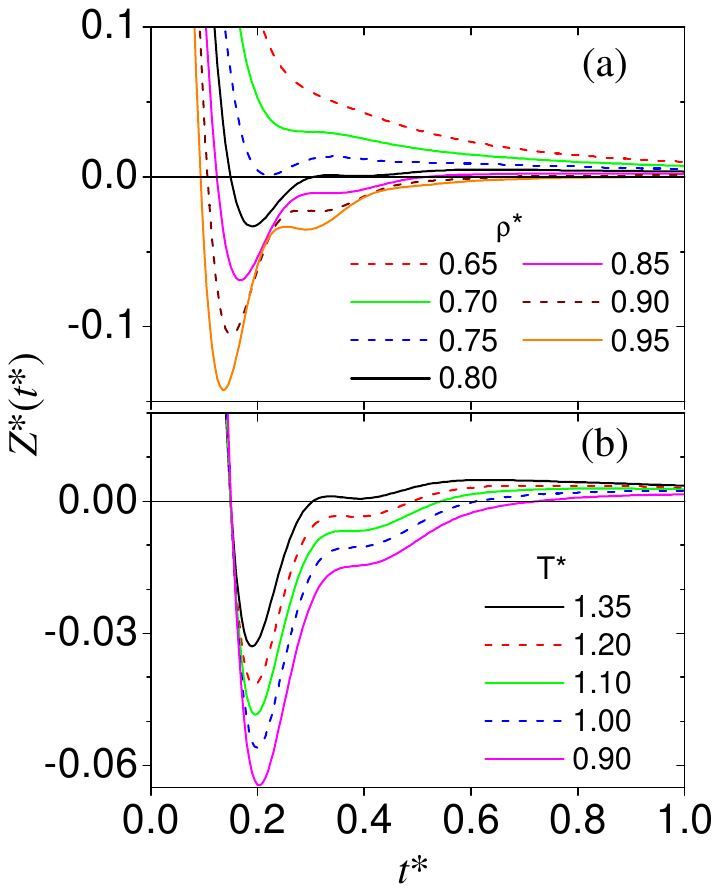}}
\caption{(Color online) Simulated VAF for $t^*\leq 1$. Only the part of $Z(t)$ close to the time axis is shown, and the steep decrease from the initial value is omitted for clarity. (a) States along the $T^*=1.35$ isotherm at the indicated densities. Density increases from top to bottom. (b) States along the $\rho^*$ = 0.80 isochore at the indicated temperatures. Temperature decreases from top to bottom. The middle line in (a) and the top line in (b) are the same curve.}
\label{fig:vaf_linear} 
\end{figure} 

All simulations were performed, and the respective VAF's computed, exactly in the same way as in the previous work \cite{paper1}, to which we refer the reader for any detail, here recalling only that the same number of particles $N=10976$ was used in a cubical box for all states, and that simulations of $10^7$ timesteps, with $\Delta t^*=0.001$, were carried out in the isokinetic ensemble and repeated ten times in order to estimate statistical uncertainties. The result is a set of normalized velocity autocorrelation functions

\begin{equation}\label{vaf_norm}
Z(t)=\frac{\langle \mathbf{v}(0)\cdot \mathbf{v}(t)\rangle}{\langle \mathbf{v}(0)^2\rangle},
\end{equation}

\noindent where $\mathbf{v}(t)$ is the velocity of a particle at time $t$ and $\langle \cdots \rangle$ includes an average over all particles.

\begin{figure*}
\resizebox{0.98\textwidth}{!}
{\includegraphics[viewport=4.5cm 18cm 16cm 24.5cm]{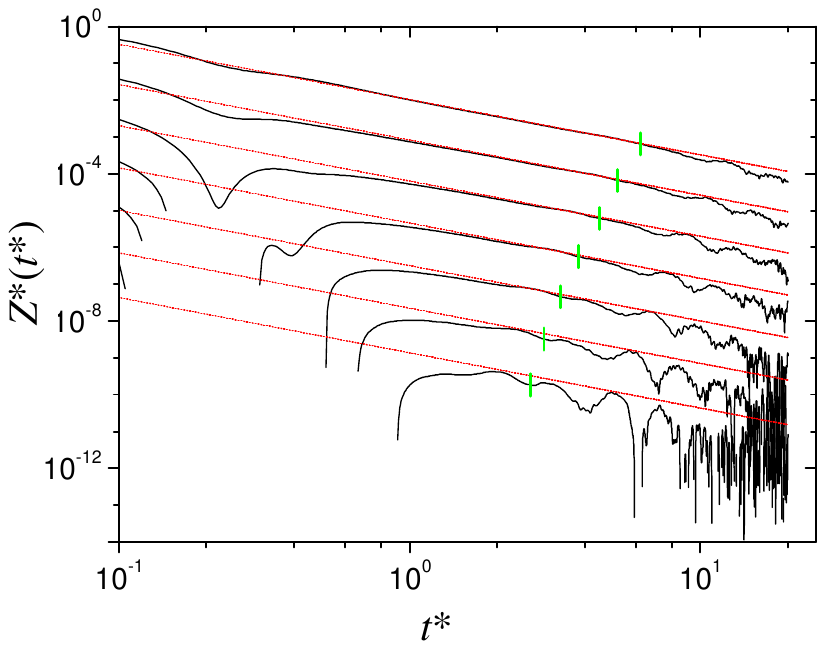}}
\caption{(Color online) Log-log plots of the VAF from simulation (black solid lines) at $t^*\geq 0.1$ for the states along the $T^*=1.35$ isotherm. The $A\,t^{-3/2}$ behavior (red straight lines) is calculated with the theoretical values of the coefficient $A$ from Eq. (\ref{eq:A-balucani}). Curves are plotted in order of increasing density from top ($\rho^*$ = 0.65) to bottom ($\rho^*$ = 0.95), and for clarity each curve is shifted downwards by a factor of 10 with respect to the preceding one. The higher density curves are interrupted in the time intervals where $Z(t)$ takes negative values. The vertical green bars mark the values of $t_{\rm R}^*$. See text for details.}
\label{fig:vaf_isoth} 
\end{figure*}

\begin{figure*}
\resizebox{0.98\textwidth}{!}
{\includegraphics[viewport=4.5cm 18cm 16cm 24.5cm]{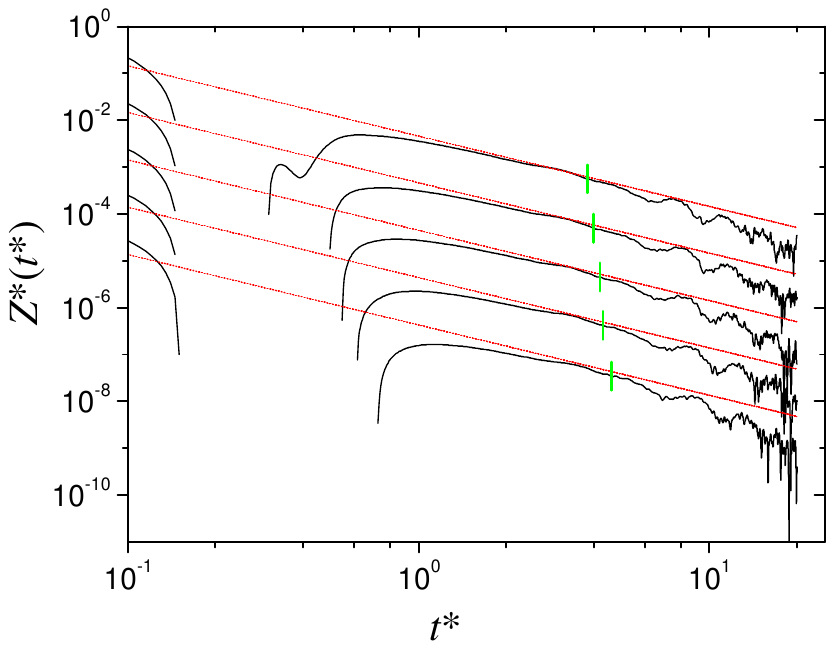}}
\caption{(Color online) Same as in Fig.\ \ref{fig:vaf_isoth} for the thermodynamic states along the $\rho^*=0.80$ isochore. Curves are plotted in order of decreasing temperature from top ($T^*$ = 1.35) to bottom ($T^*$ = 0.90), with each curve shifted downwards by a factor of 10 with respect to the preceding one. The top curve is also displayed in Fig.\ \ref{fig:vaf_isoth}.}
\label{fig:vaf_isoch} 
\end{figure*} 
For all thermodynamic states considered here, we show in Fig.\ \ref{fig:vaf_linear} the low-intensity part of $Z(t)$ at small values of the reduced time $t^*=t/\sqrt{m\sigma^2/\epsilon}$. This is where the mentioned changes in shape take place. In detail, starting from a monotonous decay of $Z(t)$, the increase of density at constant $T$ [Fig.\ \ref{fig:vaf_linear}(a)] leads to the the presence of a plateau which soon changes into a relative minimum. This already happens when $Z(t)$ still remains everywhere positive, but with a further density increase this minimum deepens and shifts to smaller times, the curve crosses the zero axis to become negative, and another shallow minimum appears at a later time. The changes of shape on cooling the system at constant density, shown in Fig.\ \ref{fig:vaf_linear}(b), are less marked and are characterized essentially by a strengthening of the negative portion of $Z(t)$.

The behavior of $Z(t)$ at long times is best appreciated in Figs.\ \ref{fig:vaf_isoth} and \ref{fig:vaf_isoch} where the log-log scale is used for an easy comparison with a power-law time dependence. As already recalled \cite{paper1}, the calculated VAF beyond a certain time is affected by spurious effects due to the use of periodic boundary conditions \cite{Allen, Hansen, Erpenbeck}. This occurs for time lags greater than the so-called recurrence time $t_{\rm R}=(N/\rho)^{1/3}/c_{\rm s}$ required by a density fluctuation to propagate over the box length at the adiabatic sound speed $c_{\rm s}$. Values of $t_{\rm R}$, calculated using sound speed data \cite{Johnson}, are reported in Tables \ref{tab:tempi_isoth} and \ref{tab:tempi_isoch}. We also display $t_{\rm R}$ in Figs.\ \ref{fig:vaf_isoth} and \ref{fig:vaf_isoch}, where the recurrence problem shows up as a spurious overall reduction of the VAF intensity, which also displays some oscillatory and rapidly increasing noisy behavior, for $t>t_{\rm R}$. It can be noted that, at constant temperature, $t_{\rm R}$ markedly decreases with increasing density due to the simultaneous reduction of the box size and increase of $c_{\rm s}$, while, at constant density, $t_{\rm R}$ increases weakly with decreasing temperature due to the reduction of $c_{\rm s}$ brought about by the lowering of the particle thermal speed. Further analysis of the VAF, described in later Sections, will be performed, at each state, using data in the respective $0 \leq t \leq t_{\rm R}$ ranges only. 

\begin{table*}
\begin{ruledtabular}
\begin{tabular}{c | c c c c c c c}
$\rho^*$ &  0.65  &  0.70  &  0.75  &  0.80  &  0.85  &  0.90  &  0.95 \\
$T^*$ &  1.35  &  1.35  &  1.35  &  1.35  &  1.35 &  1.35  &  1.35 \\
\hline
$t_{\rm R}^*$ & 6.2  & 5.2  & 4.5  & 3.8  & 3.3  &  2.9 &  2.6 \\
$\tau_{\rm E}^*$ & 0.065  & 0.054  & 0.045  & 0.038  & 0.031  & 0.026  & 0.022 \\
$t_{\rm R}^*/\tau_{\rm E}^*$ & 96 & 96 & 100 & 101 & 105 & 111 & 120 \\
$D^*$ (from Ref.\ [\onlinecite{Meier3}]) & 0.183 & 0.150 & 0.122 & 0.097 & 0.076 & 0.058 & 0.042 \\
$\eta^*$ (from Ref.\ [\onlinecite{Meier3}]) & 0.99 & 1.21 & 1.52 & 1.97 & 2.59 & 3.37 & 4.81 \\
$10^2 A^*$ (from Eq.\ (\ref{eq:A-balucani})) & 1.03 & 0.83 & 0.64 & 0.46 & 0.32 & 0.22 & 0.14 \\
\end{tabular}
\end{ruledtabular}
\caption{Recurrence time for the simulations and properties of the LJ system on the $T^*=1.35$ isotherm. The Enskog mean time between collisions for hard spheres at the same density is given, together with the diffusion coefficient and the viscosity \cite{Meier3}. The coefficient of the $A\,t^{-3/2}$ power law is obtained through Eq.\ (\ref{eq:A-balucani}). All quantities are expressed in reduced units.}
\label{tab:tempi_isoth}
\end{table*}

\begin{table*}
\begin{ruledtabular}
\begin{tabular}{c | c c c c c}
$\rho^*$ &  0.80  &  0.80  &  0.80  &  0.80  &  0.80 \\
$T^*$ &  1.35  &  1.20  &  1.10  &  1.00  &  0.90 \\
\hline
$t_{\rm R}^*$ & 3.8 & 4.0  & 4.2  & 4.3  & 4.6 \\
$\tau_{\rm E}^*$ & 0.038  & 0.040  & 0.042  & 0.044  & 0.046 \\
$t_{\rm R}^*/\tau_{\rm E}^*$ & 101 & 100 & 101 & 98 & 100 \\
$D^*$ (from Ref.\ [\onlinecite{Meier3}]) & 0.097 & 0.085 & 0.077 & 0.068 & 0.060 \\
$\eta^*$ (from Ref.\ [\onlinecite{Meier3}]) & 1.97 & 1.97 & 2.02 & 2.05 & 2.09 \\
$10^2 A^*$ (from Eq.\ (\ref{eq:A-balucani})) & 0.46 & 0.46 & 0.45 & 0.44 & 0.43 \\
\end{tabular}
\end{ruledtabular}
\caption{Same as in Table \ref{tab:tempi_isoth} for the thermodynamic states along the $\rho^*=0.80$ isochore. The first state at $T^*=1.35$ also appears in Table\ \ref{tab:tempi_isoth}.}
\label{tab:tempi_isoch}
\end{table*}

The above mentioned theories \cite{Ernst1970, Ernst1971, Dorfman} of the LTT power-law behavior give for the coefficient $A$ the formula \cite{Balucani}

\begin{equation}
\label{eq:A-balucani}
A=\frac{1}{12\rho [\pi(D+\nu)]^{3/2}},
\end{equation}

\noindent where $D$ is the self diffusion coefficient, and $\nu=\eta/(m\rho)$ and $\eta$ are the kinematic and shear viscosities, respectively. Using the literature values of $D$ and $\eta$ \cite{Meier3} reported in Tables\  \ref{tab:tempi_isoth} and \ref{tab:tempi_isoch}, we applied Eq. (\ref{eq:A-balucani}) to calculate $A$ and to draw the power-law dependence in Figs.\ \ref{fig:vaf_isoth} and \ref{fig:vaf_isoch}. It appears that at all thermodynamic states the VAF tail approaches the $A\,t^{-3/2}$ line, but such a time dependence could possibly be determined quantitatively {\it by a fit} only for $\rho^*\leq 0.75$ at $T^*=1.35$, while at higher densities and for all states at lower temperatures there is no time range where $Z(t)$ overlaps the power-law behavior. As pointed out also by McDonough et al.\ \cite{McDonough}, this fact excludes the possibility of a {\it direct} detection of the LTT behavior at high density, although in other works different claims have been made as, for example, by Meier et al.\ \cite{Meier2} who found a power-law dependence in LJ also at high density but with a different exponent, or by Williams et al.\ \cite{Williams} who did find a $t^{-3/2}$ behavior in an HS liquid. We will show in Sect.\ \ref{subsect: LTT} that the presence of an LTT can be established through the exponential mode analysis.
 
Following the procedure of Ref.\ [\onlinecite{paper1}] for consistency with the analysis there performed, we measure the characteristic times of the various dynamical processes in units of the Enskog mean free time $\tau_{\rm E}$ of a corresponding HS fluid at the same density. Enskog kinetic theory predicts a mean time between collisions given by \cite{Boon}

\begin{equation}
\label{eq:tauE}
\tau_{\rm E}=\frac{1}{4\rho\sigma^2g(\sigma)}\sqrt{\frac{m}{\pi k_{\rm B}T}},
\end{equation}

\noindent where $\sigma$ is the sphere diameter and the pair distribution function at contact $g(\sigma)$ can be obtained from the Carnahan--Starling HS equation of state as \cite{Hansen}:

\begin{equation}
\label{eq:gsigma}
g(\sigma)=\frac{1-\pi\rho\sigma^3/12}{(1-\pi\rho\sigma^3/6)^3}.
\end{equation}

\noindent The values of $\tau_{\rm E}$, obtained with the densities of the LJ states under study, are given in Tables\ \ref{tab:tempi_isoth} and \ref{tab:tempi_isoch}, where it also appears that the upper bound $t_{\rm R}$ on the useful time range corresponds to about one hundred collisions at all thermodynamic points.

\section{Exponential mode analysis}
\label{sect: multiexp}

In this Section we briefly recall the main concepts of the exponential mode analysis already applied to lower-density states [\onlinecite{paper1}] and used in the present work as well. The starting point is the recently presented \cite{Barocchi_2012, Barocchi_2013, Barocchi_2014} result that the generalized Langevin equation for a normalized autocorrelation function $C(t)$ of a classical many-body system has an exact solution written as an infinite sum of exponential functions, i.e.

\begin{equation}
\label{eq:exp_series}
C(t)=\sum_{j=1}^{\infty}I_j\exp(z_jt).
\end{equation}

\noindent Here, each term of the series can be considered as a  characteristic decay mode of $C(t)$ having $I_j$ and $z_j$ as its amplitude and frequency, respectively. If $I_j$ and $z_j$ are complex quantities, the corresponding term and its complex conjugate are both present in the series and add up to an exponentially damped oscillation. Otherwise, real $I_j$ and $z_j$ define a purely exponential decay. In all cases, ${\rm Re}\,z_j$ is negative ensuring the decay to zero of $C(t)$ in the $t\to\infty$ limit. Since $C(t)$ is an even function, one should take in the exponent of each term the absolute value $|t|$, but this can be avoided by considering positive $t$ only. Together with $C(t)$, we shall also consider its frequency spectrum $\hat{C}(\omega)$ obtained by Fourier transformation (FT):

\begin{equation}
\label{eq:FT}
\hat{C}(\omega)=\frac{1}{2\pi}\int_{-\infty}^{\infty}dt\,e^{-i\omega t}\,C(t).
\end{equation}

While the rigorous expression of $C(t)$ is given by an infinite number of terms, any practical use requires to carry out a suitable approximation by truncating the series (\ref{eq:exp_series}) after a few terms. In Ref. [\onlinecite{Barocchi_2012}] we have shown that the approximation 
introduced by the truncation consists in the neglect of higher-order derivatives of the dynamical variable, say $a(t)$, in the description of the time evolution of its autocorrelation $C(t)=\langle a(0)a(t)\rangle/\langle a^2\rangle$.

The zero-time values of $C(t)$ and of its derivatives are related to the spectral moments $\langle \omega^k\rangle = \int^{\infty}_{-\infty} d\omega\: \omega^k\: \hat{C}(\omega)$ by the relationship

\begin{equation}
\label{eq:moments}
\left(\frac{d^k C(t)}{dt^k}\right)_{t=0}=i^k\langle \omega^k\rangle.
\end{equation}

\noindent Since in a system of classical particles interacting via a continuous potential $C(t)$ has derivatives of any order at $t=0$, Eq.\ (\ref{eq:moments}) implies that $ \langle \omega^k\rangle$ is zero for all odd $k$ and has finite values for all even $k$ and, in combination with Eq.\ (\ref{eq:exp_series}), leads to an infinite set of so-called sum rules of the form

\begin{equation}
\label{eq:sum_rules}
\sum_{j=1}^{\infty} I_j z_j^k=i^k \langle \omega^k\rangle
\end{equation}

\noindent for $k\geq 0$. When the series (\ref{eq:exp_series}) is truncated, the number of applicable sum rules is also limited, as $k$ cannot exceed (though it can be less than) the number of terms in the exponential series.

\begin{table*}
\begin{ruledtabular}
\begin{tabular}{c c | c c c c c c c c}
$\rho^*$ & $T^*$ & C1 & C2 & C3 & R0 & R1 & R2  & R3 & R4  \\
\hline
$0.65$ & $1.35$ & $\times$ & $\times$ & \textemdash & \textemdash & $\times$ & $\times$  & $\times$ & $\times$ \\
$0.70$ & $1.35$ & $\times$ & $\times$ & $\times$ & \textemdash & \textemdash & $\times$  & $\times$ & $\times$ \\
$0.75$ & $1.35$ & $\times$ & $\times$ & $\times$ & \textemdash & \textemdash & \textemdash  & $\times$ & $\times$ \\
$0.80$ & $1.35$ & $\times$ & $\times$ & $\times$ & \textemdash & \textemdash & \textemdash  & $\times$ & $\times$ \\
$0.85$ & $1.35$ & $\times$ & $\times$ & $\times$ & \textemdash & \textemdash & \textemdash  & $\times$ & $\times$ \\
$0.90$ & $1.35$ & $\times$ & $\times$ & $\times$ & $\times$ & \textemdash & \textemdash  & \textemdash & $\times$ \\
$0.95$ & $1.35$ & $\times$ & $\times$ & $\times$ & $\times$ & \textemdash & \textemdash  & \textemdash & $\times$ \\
\hline
$0.80$ & $1.35$ & $\times$ & $\times$ & $\times$ & \textemdash & \textemdash & \textemdash  & $\times$ & $\times$ \\
$0.80$ & $1.20$ & $\times$ & $\times$ & $\times$ & \textemdash & \textemdash & \textemdash  & $\times$ & $\times$ \\
$0.80$ & $1.10$ & $\times$ & $\times$ & $\times$ & \textemdash & \textemdash & \textemdash  & $\times$ & $\times$ \\
$0.80$ & $1.00$ & $\times$ & $\times$ & $\times$ & \textemdash & \textemdash & \textemdash  & \textemdash & $\times$ \\
$0.80$ & $0.90$ & $\times$ & $\times$ & $\times$ & \textemdash & \textemdash & \textemdash  & \textemdash & $\times$ \\
\end{tabular}
\end{ruledtabular}
\caption{Composition of the fit models. Labels C1 to C3 denote pairs of complex conjugate modes. The modes labeled R0 to R4 correspond to real exponentials. Each mode is indicated as present ($\times$) or absent (\textemdash) in the fit model for a given thermodynamic state. The upper part of the Table refers to the $T^*=1.35$ isotherm, the lower part to the $\rho^*=0.80$ isochore.}
\label{tab:modi}
\end{table*}

Specific models for a given dynamical variable autocorrelation function $C(t)$ are then obtained by looking for a set of exponential terms that is best fitted to the data, depending on the total number of terms, the number of real and complex ones, and the set of sum rules that are chosen to be obeyed.
As already discussed in Ref.\ [\onlinecite{paper1}], there is no arbitrariness in the composition of the fit model, if the set of exponential terms retained in the truncated series is the smallest one which yields an accurate fitting, depending on the extension and accuracy of available data, without introducing an unjustified overparametrization of the fit function. As regards sum rules, besides the normalization condition ${\displaystyle \sum_j I_j=1}$ (obtained from Eq. (\ref{eq:exp_series}) for $t=0$, or from Eq.\ (\ref{eq:sum_rules}) with $k=0$), the fit parameters can also be constrained to satisfy conditions ${\displaystyle \sum_j I_j z_j^k=0}$ for a number of odd $k$ values, each of them corresponding to the requirement that $\langle \omega^{k+1}\rangle$ be finite \cite{PRE2006}.

As recalled in Sect.\ \ref{sect: intro}, the LJ lower-density VAF data of Ref.\ [\onlinecite{paper1}] were perfectly described in the whole time range up to $t_{\rm R}$ by the sum of two pairs of complex conjugate exponentials and four real exponentials. We will show in the next Section that a dynamical crossover emerging at the higher-density states of the present work will require the modification of the fit model according to precise trends, while maintaining in all cases the same number of sum rules constraints, specified by $k=0,1,3,5$. As noted in Ref.\ [\onlinecite{paper1}], these constraints both reduce the number of free fit parameters and allow for a very accurate description of the VAF short-time dependence by ensuring finiteness of its time derivatives at $t=0$ up to the sixth order. Computational details of the fitting analysis are also reported in Ref.\ [\onlinecite{paper1}].
 
\section{Results}
\label{sect: results}

\begin{figure*}
\resizebox{0.90\textwidth}{!}
{\includegraphics[viewport=4cm 10cm 17cm 24.5cm]{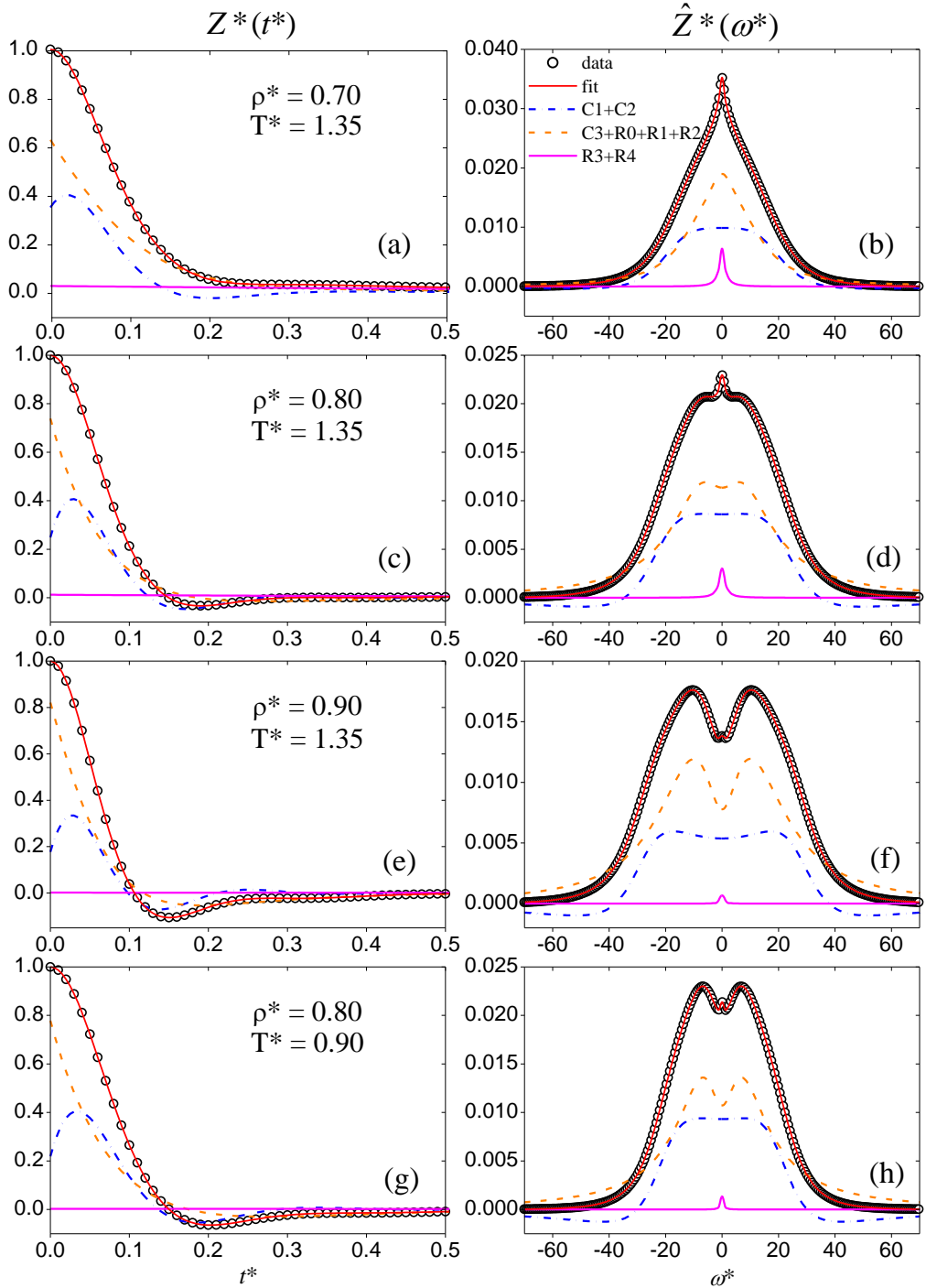}}
\caption{(Color online) $Z(t)$ and $\hat{Z}(\omega)$ at four thermodynamic states. In the left frames, MD data for the VAF (symbols) and the multi-exponential fit (red solid line through the points) are displayed at short times. The various components of the fit function, grouped together as indicated by the top right labels, are also displayed separately. In the right panels, the corresponding FT's are shown. For graphical clarity, not all available data points have been displayed.}
\label{fig:fit} 
\end{figure*}

For all thermodynamic states considered in this work, excellent fits to the VAF data in the range $0 \leq t \leq t_{\rm R}$ were obtained with the models described below and summarized in Table \ref{tab:modi}. The left frames of Fig.\ \ref{fig:fit} show examples of the fitted curves, highlighting the excellent agreement between data and models.

At the lowest density ($\rho^*=0.65$), the fit model employed in Ref.\ [\onlinecite{paper1}] works very well, in continuity with the low-density states. Keeping the same labeling of the modes, we have two pairs of complex modes (C1 and C2) and four real modes (R1 to R4, numbered in order of increasing decay time). We remind that C1 amounts to a very fast, very strongly damped oscillation, required for a good fit of the very-short-time data points, but of negligible importance at later times due to its extremely low intensity. This feature is present at all investigated states and will not be discussed further.
 
Moving along the $T^*=1.35$ isotherm towards higher densities, Table \ref{tab:modi} evidences two modifications of the optimum fit model: (a) a third pair of complex modes (C3) appears for $\rho^*\ge 0.70$, and (b) the faster-decaying real exponentials disappear (R1 first for $\rho^*\ge 0.70$, then also R2 for $\rho^*\ge 0.75$). Finally, at the two highest densities, R3 also vanishes, while another real mode labeled R0 is required. (The reasons to keep R0 distinct from other real modes will be discussed in Sect.\ \ref{subsect: low}.) Moving along the $\rho^*= 0.80$ isochore, the fit model stays unchanged while temperature is reduced, apart from a similar disappearance of R3 at the two coldest states.

In the following Subsections we will show in detail how different dynamical processes can be associated to a sensible regrouping of the various modes. These are grouped in the way indicated by the legend in Fig.\ \ref{fig:fit}(b), for reasons explained below.

The multi-exponential analysis of $Z(t)$ facilitates considerably the interpretation of the spectrum of the VAF in terms of the contributions of different dynamical processes, because Eq.\ (\ref{eq:exp_series}) immediately translates, upon FT, into a corresponding series expression for $\hat{Z}(\omega)$ where each real exponential $I_j\exp(z_jt)$ transforms into a Lorentzian line centered at $\omega=0$ and having a half width at half maximum $-z_j$, while a pair of complex exponentials gives rise to a pair of distorted Lorentzians centered at the positions $\omega=\pm |{\rm Im}\,z_j|$ with half width at half maximum $-{\rm Re}\,z_j$ \cite{PRE2006}. Depending on the amount of damping, such a pair may appear as either a doublet of lines or one unshifted bell-shaped curve. Once the various modes are determined by fitting the appropriate model to $Z(t)$ data, one automatically obtains the decomposition of $\hat{Z}(\omega)$ in terms of the corresponding centered or shifted spectral lines. This fact is exploited in the right frames of Fig.\ \ref{fig:fit} where the spectral contribution of each group of modes is displayed together with $\hat{Z}(\omega)$ and the FT of the total fit curve. 

\subsection{The LTT}
\label{subsect: LTT}

\begin{figure}
\resizebox{0.45\textwidth}{!}
{\includegraphics[viewport=6cm 19.5cm 14cm 25.5cm]{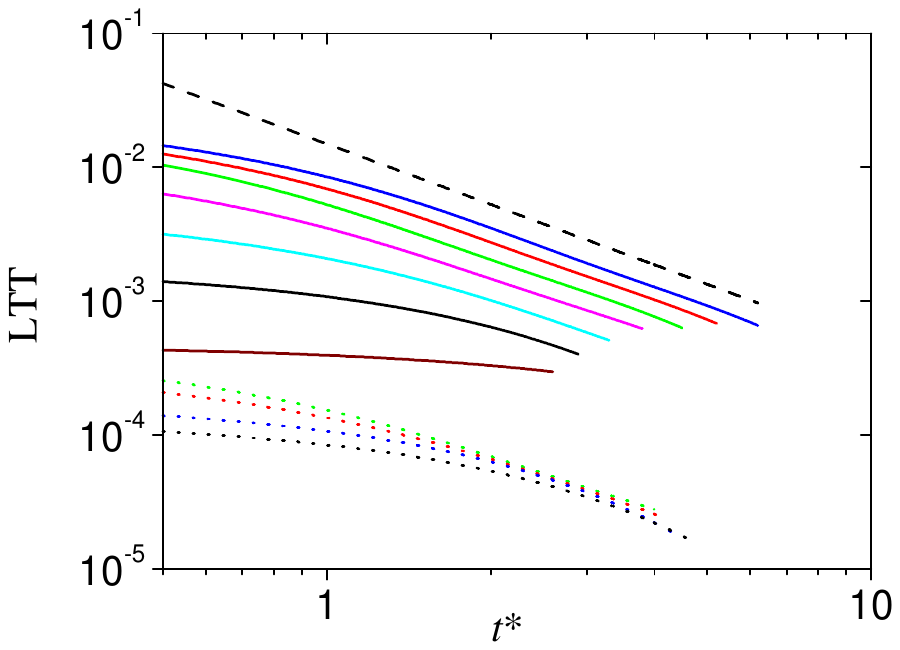}}
\caption{(Color online) Contribution of the sum of the R3 and R4 terms to the VAF time dependence. Solid curves refer to thermodynamic states at $T^*=1.35$ and are plotted in order of increasing density from top ($\rho^*$ = 0.65) to bottom ($\rho^*$ = 0.95). Dotted curves refer to states at $\rho^*=0.80$, plotted in order of decreasing temperature from top ($T^*$ = 1.20) to bottom ($T^*$ = 0.90) and have been divided by a factor of 20 for graphical clarity. All curves are plotted for $0.5\leq t^*\leq t_{\rm R}^*$. The dashed line is the $(t^*)^{-3/2}$ function, arbitrarily scaled.}
\label{fig:LTT} 
\end{figure}

At low density we showed \cite{paper1} that the R3 and, predominantly, R4 terms describe the LTT, with the latter characterized by a time constant much longer than that of the other modes and of the order of about $50-80$ if measured in units of $\tau_{\rm E}$. This fact, revealing its multi-collisional, many-particle nature in agreement with the onset of a hydrodynamiclike regime, has now been found to hold true when going towards the high-density gas or cold liquid conditions investigated here. Figure \ref{fig:LTT} displays the sum of these slowly decaying exponentials (or R4 alone when R3 is missing) and shows that an LTT is also found at the thermodynamic states studied in this work but its time dependence is compatible with an effective $t^{-3/2}$ behavior only at the lowest densities, as mentioned in Sect.\ \ref{sect: simul}.

A simple inspection of Figs.\ \ref{fig:vaf_isoth} and \ref{fig:vaf_isoch}, compared with the analogous Fig.\ 1 of Ref. \cite{paper1}, shows that the time range where an LTT can emerge as a salient feature of the VAF shrinks on the short time side due to the growth of the VAF negative part, and on the long time side because of the much smaller values of $t_{\rm R}$, beyond which the data become unreliable. This explains why at the densest or coldest states the fitting cannot determine more than just one exponential (R4). As remarked in Sect.\ \ref{sect: simul}, in so narrow a range it would be virtually impossible to fit any specific time behavior  assumed to represent directly the LTT alone. The multi-exponential analysis, however, exploiting the knowledge of the VAF in the whole time range $0 \leq t \leq t_{\rm R}$, has a far better sensitivity to, and allows for the determination of, its slowly decaying long-time part. It also follows from the above observation that a more accurate determination of the long-time VAF dependence in a dense fluid may only be obtained through a substantial increase of $t_{\rm R}$, i.e. by using larger simulation boxes with a number $N$ of atoms at least one order of magnitude larger than the present one. In such a case, the exponential expansion theory straightforwardly predicts that additional modes having even slower decay times should eventually be added to the model, although these modes would account for contributions to the VAF of negligible intensity.

The fact that the slowly decaying exponentials are indeed able to describe the LTT is even more evident if one looks at the spectra reported in the right part of Fig.\ \ref{fig:fit}, which show the presence of a tiny but clearly visible tip at $\omega=0$. This is the spectral signature of the LLT which,
assuming a $t^{-3/2}$ power law dependence, would appear as a $\sim-\sqrt\omega$ cusp \cite{Boonpage}. The multi-exponential fit shows that such a pathological frequency behavior is not justified and can be avoided, since a simple continuous function like the sum of the centered Lorentzian lines corresponding to the R3 and R4 modes accounts very well for the spectral representation of the LLT, which is displayed as the by far weakest component, confined to extremely low frequencies, of the spectra of Fig.\ \ref{fig:fit}. Accordingly, the modeling of the LTT in the form of an algebraic time dependence is also not necessary.

\subsection{Sound modes}
\label{subsect: sound}

The two complex modes forming the pair C2 define an exponentially damped oscillation that in the previous work (see Fig.\ 4 of Ref.\ \cite{paper1}) was found to be characterized by a frequency $|{\rm Im}\,z|$ close the Enskog collision frequency and a decay time $\tau=-1/{\rm Re}\,z$ slightly smaller than $\tau_{\rm E}$. As the density was increased isothermally from $\rho^*= 0.30$ to $\rho^*= 0.60$, a steady increase of $\tau /  \tau_{\rm E}$ by nearly a factor of 4, while $\omega\tau_{\rm E}$ remained practically constant, showed that this oscillatory motion becomes better defined, though still strongly damped. Also, we noted that the fractional contribution of C2 to the VAF time integral clearly increases with density (see Table II of Ref.\ \cite{paper1}) while the C1 pair contributes negligibly.

In Ref.\ \cite{paper1} we did not associate the C2 oscillation with any specific dynamical process, apart from noting that its growing intensity points to an increasing relevance of vibratory motions likely related to the bouncing of atoms off their neighbors. However, the above mentioned characteristics of this component of the VAF are all compatible with the suggestion that what is accounted for by the C2 pair is the dynamics of collective motions due to propagating longitudinal acoustic waves. In fact, sound propagation occurs at any density, but when approaching dense fluid conditions the sound speed increases and the visibility of sound modes is enhanced.

\begin{figure*}
\resizebox{0.85\textwidth}{!}
{\includegraphics[viewport=4cm 11cm 18.5cm 24.5cm]{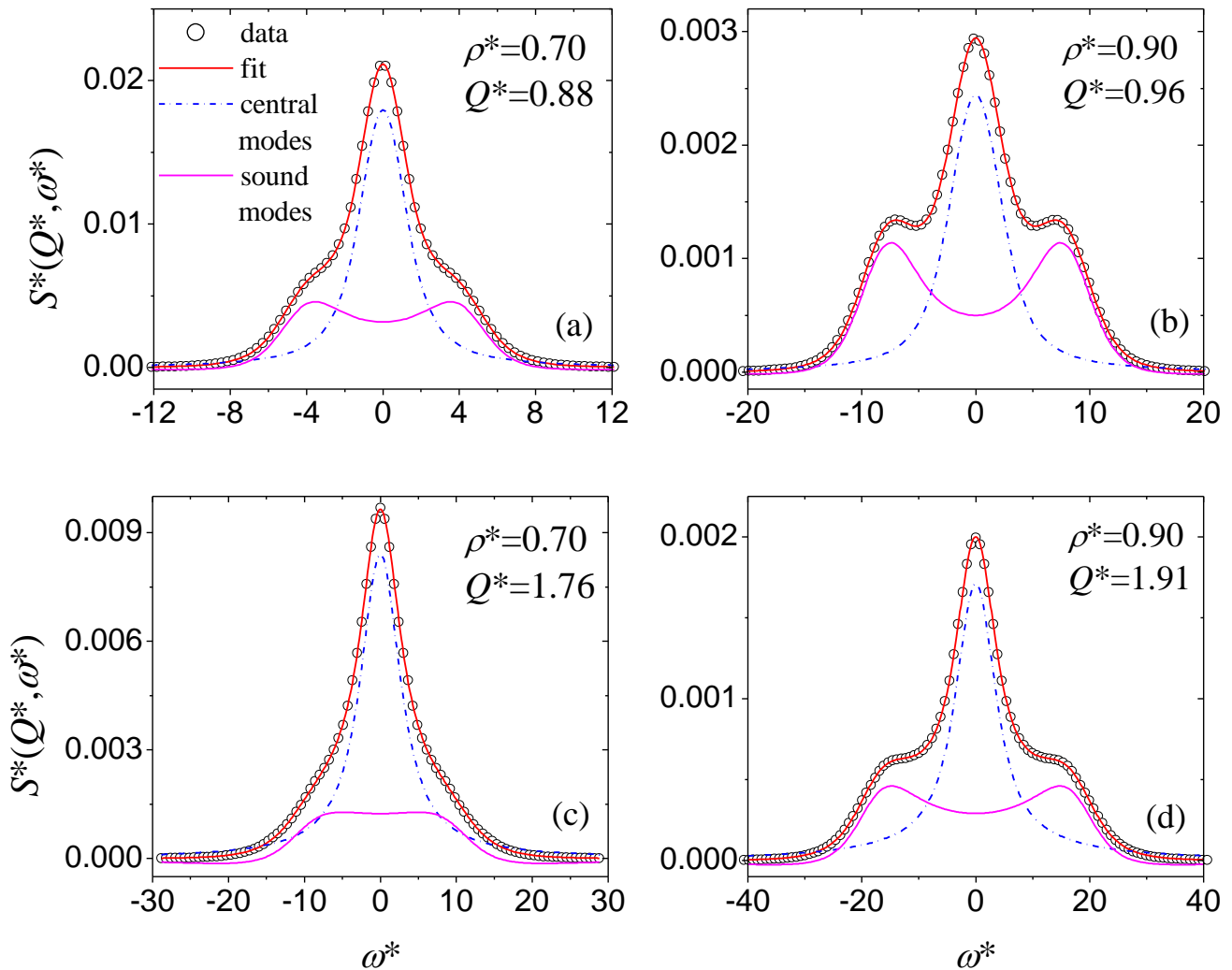}}
\caption{(Color online) Simulated $S(Q,\omega)$ (dots) at two densities and two $Q$ values indicated by the labels. Also shown are the VE fits (solid lines through the data points) and the partial contributions to the fit of the sum of the two central modes and of the two sound modes.}
\label{fig:sqw} 
\end{figure*} 

The connection between VAF and collective dynamics of fluids is realized through the interpretation of the VAF frequency spectrum as a generalized overall ``density of states''. This concept is of standard use in the phonon theory of harmonic crystals and has also been applied to the anharmonic case \cite{Glyde}, to vibrations in amorphous solids \cite{Rahman} and glasses, and to internal degrees of freedom of polyatomic molecules. In the case of liquids, a connection between the frequency density distribution of vibrational states and its relation to the spectrum of the VAF has been proposed in the framework of the so-called normal mode analysis in various formulations \cite{Garberoglio, Cao}. It has to be noted that the relationship between collective dynamics and density of states is a general property, not restricted to the longitudinal wave case only. For example, in a recent work \cite{Zanatta}, the density of states of the crystal phase has been used to interpret dynamic structure factor data of a liquid metal, where an excitation was observed at frequencies close to those of the transverse phonon density of states and was therefore assigned a transverselike character.

As already stated in Sect.\ \ref{sect: intro}, the VAF and its spectrum contain all kinds of dynamical information relative to both diffusive and vibrational motions. This was evidenced by the treatment of Gaskell and Miller \cite{Gaskell1} where, under reasonable assumptions, $Z(t)$ could be written in terms of both self and collective functions. Here we show, in particular, that the C2 pair has a clear and direct quantitative connection to the dispersion curve of the longitudinal acoustic modes.

In order to do so, we also carried out MD simulations of the intermediate scattering function $F(Q,t)$ and, through Fourier transform, of the dynamic
structure factor $S(Q,\omega)$ at a number of wavevectors $Q$ for each density. In this case we used 2048 particles with a simulation length of $1.2\cdot10^6$ timesteps of duration $\Delta t^*=0.001$, and the $Q$ values were multiples of the minimum $Q$ value allowed, at each density, by the respective box size. With the same simulations we also computed the transverse current correlation spectra that will be discussed later in Sect.\ \ref{subsect: low}. A smaller number of particles than in the simulations of the VAF was possible because in this case the correlation functions to be computed decay faster, without long-time tails, and a shorter $t_{\rm R}$ is sufficient, allowing for smaller simulation boxes.

\begin{figure*}
\resizebox{0.98\textwidth}{!}
{\includegraphics[viewport=4cm 17cm 17cm 25cm]{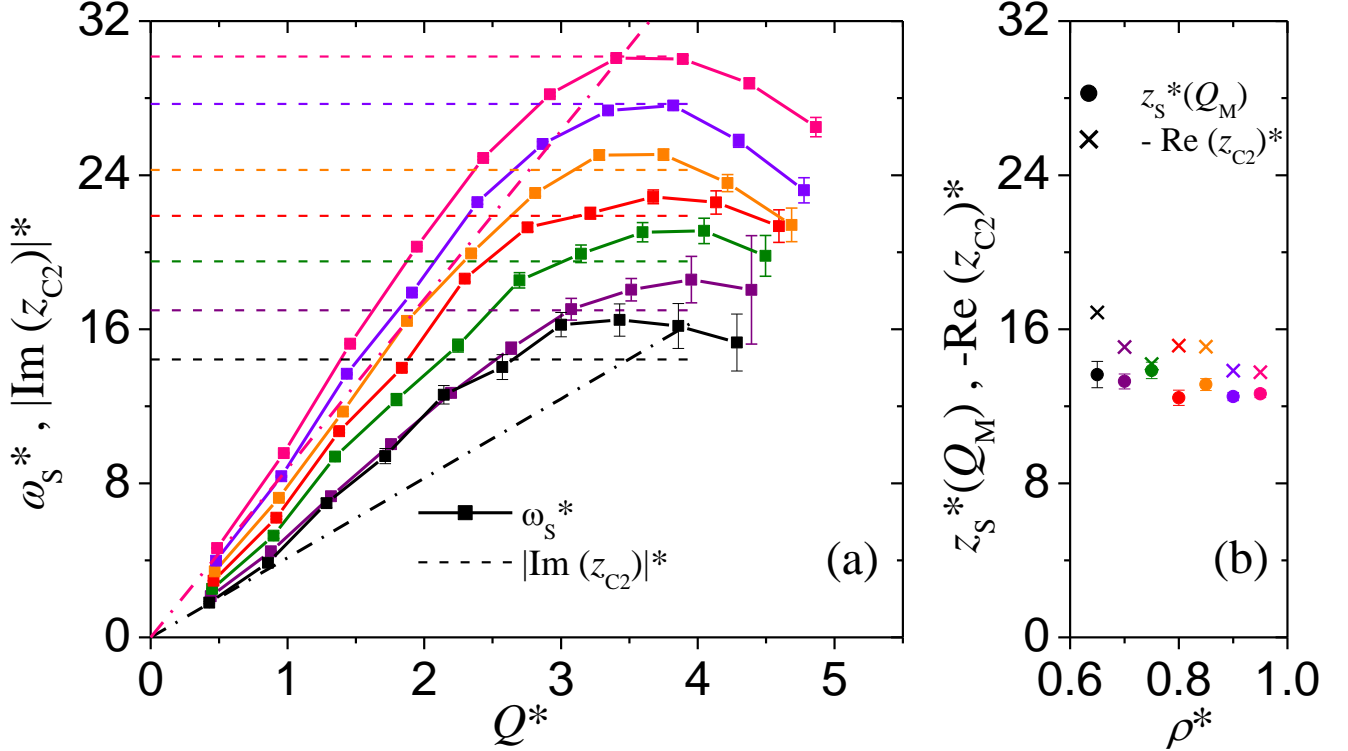}}
\caption{(Color online) (a) The curves joining symbols with error bars are the dispersion curves $\omega^*_{\rm s}(Q^*)$ of the longitudinal acoustic mode along the $T^*=1.35$ isotherm, in order of increasing density from bottom ($\rho^*=0.65$) to top ($\rho^*=0.95$). Dashed lines, in the same order, mark the corresponding values of the C2 oscillation frequency. The hydrodynamic dispersion straight lines $c_{\rm s}^*Q^*$ are also shown (dash-dots) for the highest and the lowest density. (b) Damping $z^*_{\rm s}(Q_{\rm m}^*)$ of the longitudinal acoustic mode (dots with error bar) at $Q^*=Q_{\rm m}^*$, i.e. where the respective dispersion curve reaches its maximum value in (a), and damping $-{\rm Re}\,(z_{\rm C2})^*$ of the complex pair C2 (crosses). The color code is the same in the two frames.}
\label{fig:sound1} 
\end{figure*} 

\begin{figure}
\resizebox{0.50\textwidth}{!}
{\includegraphics[viewport=6cm 17cm 16cm 25cm]{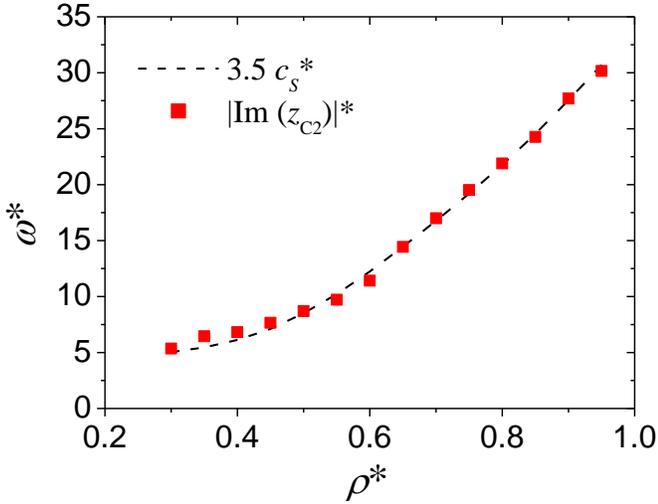}}
\caption{(Color online) Oscillation frequency of the C2 exponential (full squares) as a function of density at $T^*=1.35$ including the low-density states of Ref.\ \cite{paper1}. The dashed line depicts the $3.5\,c_{\rm s}^*$ behavior explained in the text.}
\label{fig:sound2} 
\end{figure}
 
We analyzed the $S(Q,\omega)$ spectra by fitting to them the viscoelastic (VE) model line shape, which has proven to ensure a very accurate description of the dynamics of density fluctuations in a very large variety of fluids \cite{Guarini}. The main properties and the expression of the VE model are briefly summarized in the Appendix. Examples of $S(Q,\omega)$ simulated spectra with the corresponding fitted ones are shown in Fig.\ \ref{fig:sqw}. The fits are of very good quality at all $Q$ and density values, and allowed us to extract the parameters relevant to the description of sound modes, namely the frequency $\omega_{\rm s}$ and the damping $z_{\rm s}$, which are plotted in Fig.\ \ref{fig:sound1}(a), displaying the typical shape of the sound dispersion curves of a fluid, and Fig.\ \ref{fig:sound1}(b), respectively.  

In much the same way as the density of phonon states in crystals peaks at frequencies where the dispersion branches have a horizontal tangent \cite{Ashcroft}, the contribution of sound modes to the total VAF spectrum of a liquid is also expected to be centered around the top value of the sound dispersion curve. Figure \ref{fig:sound1}(a) shows that the frequency of the fitted C2 pair is very close to the maximum sound frequency, with an increasingly better agreement for increasing density. In Fig.\ \ref{fig:fit}, contrary to the crystal case, $\hat{Z}(\omega)$ does not display a clear shoulder at the acoustic frequencies ${\pm|\rm Im}\,z_{\rm C2}|$, because its overall shape is also determined by other contributions of different dynamical origin. Here one can well appreciate the advantage brought about by the possibility of extracting specific partial contributions to the total $\hat{Z}(\omega)$. The spectral partial component formed by the C1 and C2 mode group does, in fact, show clear shoulders at nonzero frequencies, as seen in Fig.\ \ref{fig:fit}.  

Besides the frequency, also the broadening of the acoustic line in $S(Q,\omega)$ is directly reflected by a corresponding broadening of the density of states. Accordingly, Fig.\ \ref{fig:sound1}(b) shows that the damping $-{\rm Re}\,z_{\rm C2}$ of the C2 exponentials matches very closely the damping parameter $z_{\rm s}(Q_{\rm m})$ of the acoustic spectral lines if $Q$ is taken to be the position $Q_{\rm m}$ of the maximum in the dispersion curve. This is in agreement with the already stated fact that the density of states is dominated by the sound modes of maximal frequency.

It is worth noting from Fig.\ \ref{fig:sound1} that the top sound frequency varies with density between reduced values of about 16 and 30, while the corresponding dampings stay constant around 13, indicating that the acoustic excitation becomes less sharply defined in less dense fluids. A consequence of this is that also sound modes of frequency not much lower than the maximum one can contribute partially to the density of states of a fluid, the more so the lower the density. This fact explains why ${|\rm Im}\,z_{\rm C2}|$ becomes smaller than the maximum of $\omega_{\rm s}(Q)$ when density is decreased. Analogously, the contribution of lower-frequency sound modes with their respective broadenings makes the damping of the C2 modes a bit larger than the damping $z_{\rm s}$ of the acoustic excitation at the top of the dispersione curve, as shown in Fig.\ \ref{fig:sound1}(b).

The strict relation between the oscillation frequency of the C2 modes and the collective longitudinal dynamics can also be visualized in another way. In Fig.\ \ref{fig:sound2}, ${|\rm Im}\,z_{\rm C2}^*|$ is plotted as a function of density, including for completeness the lower density states of Ref.\ \cite{paper1}. The line is the plot of $3.5\, c_{\rm s}^*$, which describes very well the density behavior of the data points. Here $3.5$ is only an empirically adjusted number, to which no specific meaning is attached. However, there are reasons to justify this kind of relationship and a numerical factor close to and slightly larger than $\pi$. Figure \ref{fig:sound1}(a) shows that, for increasing $Q^*$, the dispersion curves bend upwards (positive dispersion) before reaching their top values, and it is also found that they intersect the hydrodynamic linear dispersion $c_{\rm s}^*Q^*$ somewhere near the position $Q_{\rm m}^*$ of the maximum. Beyond this point, dispersion curves in fluids drop typically a bit faster than they rise to the left of it, up to the value of $Q^*=Q_{\rm p}^*$ where the static structure factor $S(Q^*)$ has its main peak. Thus, $Q_{\rm m}^*$ is usually close to, but slightly greater than, $Q_{\rm p}^*/2$ and, analogously, the maximum frequency is a bit higher than $c_{\rm s}^*Q_{\rm p}^*/2$. On the other hand, in a fluid $Q_{\rm p}^*$ is close to $2\pi$, therefore the top frequency of the acoustic mode is well approximated by the hydrodynamic dispersion evaluated at $Q^*\gtrsim \pi$, which gives the result displayed in Fig.\ \ref{fig:sound2}.

As pointed out above in the discussion of Fig.\ \ref{fig:sound1}, with decreasing density $|{\rm Im}\,z_{\rm C2}^*|$ becomes lower than the top value of the dispersion curve, but also $Q_{\rm p}^*$ slightly decreases below the value $2\pi$ due to a less closely packed structure of the fluid. These two weak variations appear to effectively compensate each other so that the relation $|{\rm Im}\,z_{\rm C2}^*|\approx 3.5\, c_{\rm s}^*$ continues to hold at low densities as well.

While the above is just a semi-quantitative argument, the agreement displayed in Fig.\ \ref{fig:sound2} remains remarkable, highlighting a simple but accurate proportionality relation between two properties of the system under study: one, $|{\rm Im}\,z_{\rm C2}|$, derived from the theoretically well-founded exponential mode expansion, the other, $c_{\rm s}$, being a fundamental thermophysical quantity of a fluid. The link between these two properties is meaningful independently of possible minor changes of the actual proportionality factor.

In this Subsection, we have thus definitely established in a quantitative way that the exponential expansion of the VAF allows to extract the part of the density of states that is strictly and directly connected to the dynamics of longitudinal sound wave propagation. We remark that, in doing so, we relate the dispersion curves and the VAF spectrum \emph{of the same fluid state}, without using any information on the vibrational density of states of the corresponding crystalline solid. In this respect, our findings conform to the results of the Gaskell and Miller \cite{Gaskell1} approach that explicitly links $Z(t)$ to the self and collective correlation functions of one and the same system. However, our method deepens considerably the analysis by providing a means to separate different dynamical processes, opening up the way for studying their evolutions as functions of any relevant state parameters.

\subsection{Low-frequency dynamics}
\label{subsect: low}

So far we have identified the VAF expansion modes related to the LTT (i.e. R3 and R4) and the longitudinal collective dynamics (the C2 pair, remembering that the contribution of the C1 pair is by far negligible). We now turn to the remaining modes. As shown in Table \ref{tab:modi}, these include the real exponentials R1 and R2 which can be qualified as fast decaying terms, since their decay times are of the order of $\tau_{\rm E}$. (We find $\tau_{\rm R1}/ \tau_{\rm E}=0.86$ for $\rho^*=0.65$ and $\tau_{\rm R2}/ \tau_{\rm E}=3.30$ and 3.48 for $\rho^*=0.65$ and 0.70, respectively, where the decay time for a real mode $j$ is defined as $\tau_j=-1/z_j$.)

\begin{figure*}
\resizebox{0.85\textwidth}{!}
{\includegraphics[viewport=4cm 18.5cm 17.5cm 25cm]{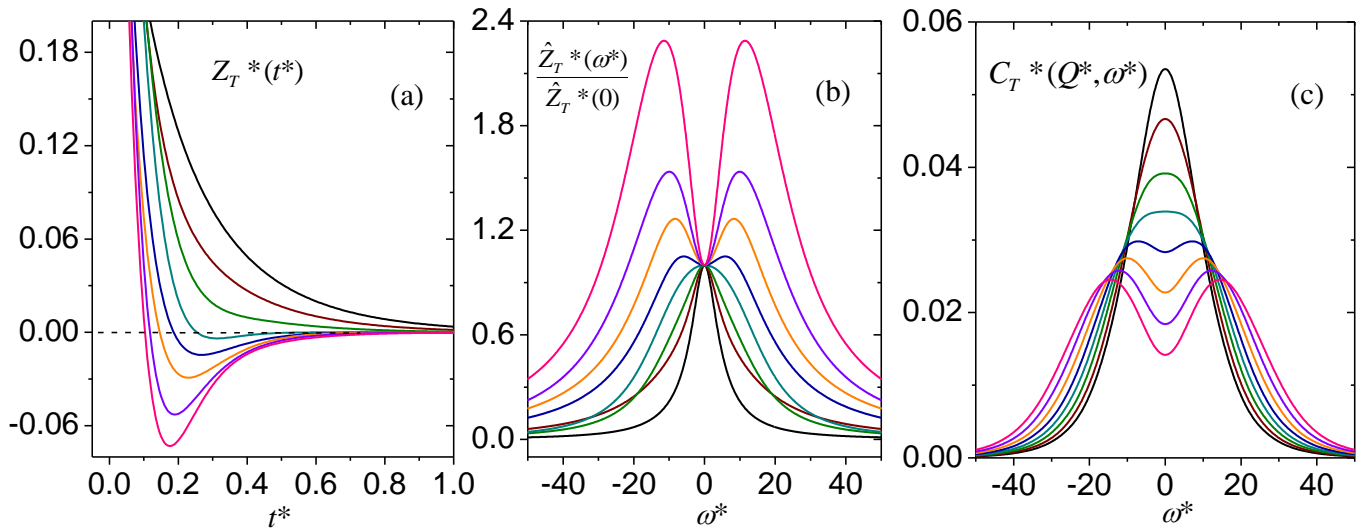}}
\caption{(Color online) (a) Time dependence of $Z_{\rm T}(t)$, i.e. sum of the fitted exponentials belonging to the low-frequency group. Curves are ordered for increasing density from top ($\rho^*=0.60$) to bottom ($\rho^*=0.95$) in steps of $0.05$. (b) Frequency spectra of the curves in (a), normalized to their respective zero frequency value. Here lower densities correspond to lower curves (at small frequency). (c) Frequency spectra of the transverse current autocorrelation function at a selected $Q$ value (see text) and for the same densities as in (a), (b). Curves are ordered for increasing density from highest to lowest peak height. The color code is the same in all three frames.}
\label{fig:transverse} 
\end{figure*}

However, we noted that optimization of the fit model leads to the elimination of both terms while introducing a new complex pair C3 already at density $\rho^*=0.70$. This suggests that we are witnessing a transition from nonpropagating to propagating modes, with the transformation of purely decaying term(s) into another oscillatory component of the VAF time dependence. We also noted that at the highest densities a good fit requires an extra real mode (labeled R0) which turns out to have a negative amplitude. The interplay between these components of the VAF can be complicated to follow in extreme detail. However, what we are interested in here is to see how a contribution to the VAF that starts at low density as a sum of diffusive relaxation channels \cite{paper1} evolves towards an oscillatory behavior, and what kind of dynamics it corresponds to. Thus, we define another group by summing together the exponentials R1, R2, C3, and R0, whenever present. The time function so obtained, which in general is not normalized to unity at $t=0$, and its FT are denoted as $Z_{\rm T}(t)$ and $\hat{Z}_{\rm T}(\omega)$, respectively.

Both these quantities are displayed in Fig.\ \ref{fig:fit}, from which a few facts can be immediately noticed. First, $Z_{\rm T}(t)$ constitutes the largest component of the total VAF at low densities, while providing an intensity comparable to the high-frequency group (C1+C2) at high density. (See also the discussion of Table \ref{tab:integrals} in Sect.\ \ref{sect: concl}.) Second, $\hat{Z}_{\rm T}(\omega)$ shows clearly the presence of a peak at a nonzero frequency for $\rho^*\geq 0.80$. On the other hand, Table \ref{tab:modi} indicates that the C3 complex exponential pair is still missing at $\rho^*= 0.65$ but is present at $\rho^*= 0.70$. Thus a vibrational dynamics seems to set on in the crossover range $0.65<\rho^*<0.75$. In Fig.\ \ref{fig:fit}(b) (i.e. at $\rho^*= 0.70$) one would at first sight conclude that no such vibration frequency is present. However, we already noted that the spectral signature of a strongly damped oscillation may appear as a single line. Moreover, at $\rho^*= 0.70$ the C3 pair of lines is still added to a central Lorentzian line due the real exponential R2 which can mask partially the spectrum of the oscillating terms. At all densities, however, the frequency of the C3 modes is clearly smaller than that of the high-frequency group, and for this reason we shall refer to these modes as the low-frequency group, in accordance with the title of this Subsection.

In order to tentatively assign this group of modes to a specific dynamical property of the fluid, we first note that in the Gaskell and Miller approach both longitudinal and transverse dynamics appear, on equal footing, in the integral representation of the VAF \cite{Gaskell1}, but that we have not yet related any fit components of either the VAF or its FT to the dynamics of transverse modes. Moreover, it is known that if excitations in the transverse current autocorrelation function take the form of propagating waves, they do it with lower frequencies than in the longitudinal case \cite{Gaskell2,Sampoli}. Also, it is a common belief that a low-density fluid does not sustain propagation of shear waves while non-ideal dense liquids do \cite{Balucani, Brazhkin_2013, Cunsolo_2012}. All these facts point then to a likely connection of the whole low-frequency group of modes to transverse dynamics, and it appears reasonable to check if more quantitative arguments can be put forward to make this relation stand on firm grounds.

Here we go into details of this analysis for the $T^*=1.35$ states only, but analogous observations can be made for those along the $\rho^*=0.80$ isochore. The results are summarized in Fig.\ \ref{fig:transverse}, where for comparison we have also included the $\rho^*=0.60$ case. The time dependence of $Z_{\rm T}(t)$ is shown in Fig.\ \ref{fig:transverse}(a). This quantity begins to feature a negative part for $\rho^*=0.75$ that progressively deepens and shifts its minimum to lower times. Such a behavior describes the emergence of an oscillation of growing strength and increasing frequency.

It is instructive to compare the plot of $Z_{\rm T}(t)$ with that of the total VAF shown in Fig.\ \ref{fig:vaf_linear}(a), in two respects. First, the onset of a negative part of $Z(t)$ is also located just above $\rho^*=0.75$. The near coincidence of this threshold density in $Z(t)$ and $Z_{\rm T}(t)$ is remarkable, since the former contains other terms which are absent in the latter. This suggests that the VAF develops its negative part mainly because of the overall oscillating behavior of the low-frequency group (which in turn is mostly due to the C3 complex pair). Second, the next oscillation clearly visible in Fig.\ \ref{fig:vaf_linear}(a) around $t^*=0.35$ is missing in $Z_{\rm T}(t)$ and is therefore due to the C2 complex pair.

\begin{figure}
\resizebox{0.55\textwidth}{!}
{\includegraphics[viewport=5.5cm 17cm 17cm 24.5cm]{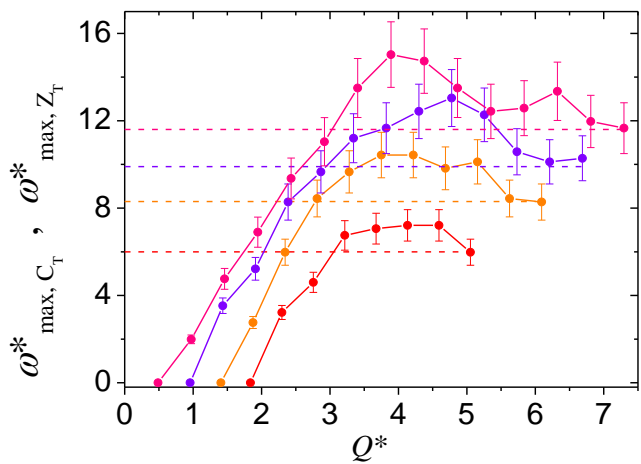}}
\caption{(Color online). The lines joining symbols with error bars are the transverse dispersion curves obtained as the positions of maxima of  $C_{\rm T}(Q,\omega)$. Data for the $T^*=1.35$ states are shown in order of increasing density from bottom ($\rho^*=0.80$) to top ($\rho^*=0.95$). Dashed lines, in the same order, mark the corresponding position of maxima of $\hat{Z}_{\rm T}(\omega)$.}
\label{fig:trans_disp} 
\end{figure}

Going to the frequency domain, $\hat{Z}_{\rm T}(\omega)$ is plotted in Fig.\ \ref{fig:transverse}(b) after normalization to its zero frequency value. This graphical representation highlights the change in shape brought about by the density increase: the curve is bell-shaped up to $\rho^*\approx 0.70$ and shows a distinct double peak for $\rho^*\geq 0.80$. At $\rho^*=0.75$ no peaks are visible but the curve features a flat top that suggests the presence of two strongly overlapping lines, so that this density can again be considered to be a threshold value, in agreement with what has already been discussed.

The connection to transverse dynamics is obtained when Fig.\ \ref{fig:transverse}(c) is considered, where we display the spectrum $C_{\rm T}(Q,\omega)$ of the transverse current autocorrelation function also obtained from the MD simulations described in Sect.\ \ref{subsect: sound} \cite{currents}. The curves refer to the various densities and to wavevector $Q^*$ values ranging from 3.1 to 3.9. (Here $Q^*$ varies slightly with density because it was taken to be a fixed multiple of the minimum value accessible in the simulation box.) $C_{\rm T}(Q,\omega)$ and $\hat{Z}_{\rm T}(\omega)$ are conceptually different quantities that it makes no sense to compare directly. Nevertheless, they are linked by a common underlying set of dynamical processes. Actually, and more precisely, for the same reasons discussed in Sect.\ \ref{subsect: sound} we can say that the latter represents the density of states related to the $Q$-dependent collective dynamics described by the former. A meaningful comparison between the spectra in Figs.\ \ref{fig:transverse}(b) and \ref{fig:transverse}(c) can be performed at the level of their evolution with density. Doing so, one sees that the overall shape of transverse current spectra and its density dependence follows closely the very similar pattern of $\hat{Z}_{\rm T}(\omega)$. Again, a flat top at $\rho^*=0.75$ points at the presence of two peaks not separated because of their width, which becomes apparent at $\rho^*=0.80$. The case $\rho^*=0.70$ cannot be clearly assigned to either the presence or the absence of propagating modes by simple visual inspection, and only the fitting of appropriate models to $C_{\rm T}(Q,\omega)$ can solve the issue. This task is outside the scope of the present paper, and here we limit ourselves to compare in Fig.\ \ref{fig:trans_disp} the frequencies of the main spectral features of $\hat{Z}_{\rm T}(\omega)$ and of $C_{\rm T}(Q,\omega)$ obtained by estimating both from the positions $\omega_{{\rm max},Z_{\rm T}}$ and $\omega_{{\rm max},C_{\rm T}}$ of their respective maxima. Though this is no rigorous procedure, a clear similarity appears to the sound mode case presented in Fig.\ \ref{fig:sound1}(a).

We have, therefore, obtained clear evidence of the link between the low-frequency mode group, extracted from the exponential expansion of the total $Z(t)$, and the dynamics of transverse collective motions in the fluid, and the subscript T used to denote this component of the total fit model for the VAF does indeed rightly assume the meaning of ``transverse''.

\subsection{Dynamical crossover}
\label{subsect: crossover}

In Sect.\ \ref{subsect: low} we have noted how the low-frequency dynamical processes that show up in the exponential mode expansion of the VAF reveal the appearance of a dynamical transition, essentially driven by the trasformation of transverse waves from nonpropagating to propagating ones.

A precise determination of the crossover point, besides requiring a very time-consuming finer exploration of the thermodynamic state space, is somewhat hindered by the diversity of criteria that may be applied to define the relevant property. For example, Table \ref{tab:modi} shows that the existence of a low-frequency oscillating component of the VAF is already attained at $\rho^*=0.70$, which is also the density where the total $Z(t)$ has an  inflection point [see Fig.\ \ref{fig:vaf_linear}(a)]. Then, the appearance of the inflection point and of the negative part of $Z(t)$ do not occur at exactly the same density. On the other hand, at $\rho^*=0.70$, besides the oscillatory part, the transverse dynamics contains also a further real mode (R2), and becomes purely oscillatory, though damped, only at $\rho^*=0.75$. In agreement with that, looking at Fig.\ \ref{fig:transverse} we have suggested a possible threshold density $\rho^*=0.75$, though only based on visual inspection. 

However, rather than trying to establish a rigid criterion for an exact determination of the crossover density, what matters here is the evidence of a dynamical transition which, for the slightly supercritical states investigated here, can be found at a density which is, broadly speaking, between those of the critical and triple points but closer to the latter. This transition separates lower density fluid states, where the dynamics has a typical gaslike character, from higher desnity ones where it takes up a more distinctly liquidlike nature.

To clarify this point, it is useful to note that, if $\hat{Z}_{\rm T}(\omega)$ is characterized by a spectral feature of nonzero frequency, and given its meaning of a partial density of states, this implies the existence of a dispersion curve of propagating (i.e. having nonzero excitation energies) transverse waves. Actually, we can go further by expecting that $\hat{Z}_{\rm T}(\omega)$ is a spectral distribution peaked around the frequency at which the transverse dispersion curve shows a flat $Q$ dependence or attains its maximum value.

On the other hand, it is well known that a dilute fluid does not support the propagation of shear waves. In quantitative terms, this means that no strictly positive frequency can be introduced, and no dispersion curve can be defined. Accordingly, the whole transverse dynamics reduces to the superposition of diffusive motions and no oscillatory behavior appears in autocorrelation functions (or parts thereof, as in the case of the VAF) related to transversal motions.

\begin{table*}
\begin{ruledtabular}
\begin{tabular}{c c | c c c }
$\rho^*$ & $T^*$ & C1$+$C2 & C3$+$R0$+$R1$+$R2 & R3$+$R4 \\
\hline
$0.65$ & $1.35$ & $26.3$ & $52.5$ & $21.3$ \\
$0.70$ & $1.35$ & $26.9$ & $51.8$ & $21.3$ \\
$0.75$ & $1.35$ & $24.7$ & $54.5$ & $20.7$ \\
$0.80$ & $1.35$ & $35.8$ & $47.3$ & $16.9$ \\
$0.85$ & $1.35$ & $43.4$ & $44.1$ & $12.4$ \\
$0.90$ & $1.35$ & $37.7$ & $54.4$ & $7.8$ \\
$0.95$ & $1.35$ & $49.8$ & $42.4$ & $7.9$ \\
\hline
$0.80$ & $1.35$ & $35.8$ & $47.3$ & $16.9$ \\
$0.80$ & $1.20$ & $38.1$ & $46.8$ & $15.0$ \\
$0.80$ & $1.10$ & $39.5$ & $47.6$ & $13.0$ \\
$0.80$ & $1.00$ & $41.2$ & $48.9$ & $9.8$ \\
$0.80$ & $0.90$ & $42.6$ & $48.9$ & $8.6$ \\
\end{tabular}
\end{ruledtabular}
\caption{Percent contribution to the integral of the VAF of the fitted modes grouped as indicated. (Missing modes in a given fit model do not contribute.) The numbers in each row add up to 100 percent to within rounding of the last digit. The upper part of the Table refers to the $T^*=1.35$ isotherm, the lower part to the $\rho^*=0.80$ isochore.}
\label{tab:integrals}
\end{table*}

Another important point is that propagation of acoustic modes, either longitudinal or transverse, depends not only on the thermodynamic properties of the fluid but also on the wavevector values, a fact critically relevant, in particular, for the transverse excitations. Indeed, it is customarily stated that a dense fluid can sustain transverse waves, but only for wavevectors above a certain value \cite{Hansen}, in agreement with the fact that in the hydrodynamic limit $Q\to 0$ the transverse current autocorrelation spectrum has the Lorentzian shape typical of diffusive processes \cite{Balucani} without any nonzero-frequency excitations, and this theoretical result holds true even for dense liquids. It follows that a transverse dispersion curve can be defined only for $Q$ larger than the minimum required. Therefore, the crossover we are dealing with is the transition between the situation where no transverse modes propagate \emph{at any} $Q$, and the one in which they propagate \emph{at least in some} $Q$ range. Consideration of only one of the two relevant variables ($Q$, $\rho$) will only allow for an incomplete account of the transverse dynamics. An evidence of this fact is provided by Fig.\ \ref{fig:trans_disp}, where the transverse dispersion curve for a lower density state begins at a larger $Q$ value. Analogous results have been very recently obtained in another study of LJ dynamics \cite{Fomin_2016}.

As recalled in Sect.\ \ref{sect: intro}, the issue of a separation between two different regions of the thermodynamic state space of a supercritical fluid, marked by a crossover boundary, has been a largely debated topic in the last years. Following seminal insight and theoretical developments of Frenkel \cite{Frenkel}, the crossover Frenkel line has been identified through the occurrence of specific properties \cite{Brazhkin_2012}, actually not all coinciding as far as the line position in the pressure-temperature plane is concerned. Among such properties are the onset of propagation of transverse waves, the emergence of positive dispersion in the propagation of longitudinal sound modes, and the temperature dependence of the constant-volume specific heat per atom crossing the value $c_V=2k_{\rm B}$. Another criterion for the definition of the Frenkel line specifically involves a property of the VAF, as it is formulated by assuming the crossover to occur when the VAF shape begins to display a relative minimum and a relative maximum instead of a monotonic decay \cite{Brazhkin_2013}. As seen in Sect.\ \ref{subsect: low}, we obtain a very similar localization of our crossover condition. However, it should be remarked that the discussion about boundary lines in the fluid phase diagram has been mostly concerned with the transition induced by changing temperature and pressure in wide ranges, including consideration of highly supercritical fluid states at very large pressures. In constrast, we have obtained evidence of a crossover driven by the density alone in isothermal conditions at temperatures just above the critical point.

\section{Conclusions}
\label{sect: concl}

The extension of the exponential mode expansion of the VAF to a wider density range than that already investigated \cite{paper1} has allowed us to follow in detail the transition between a dilute and a very dense LJ fluid at a weakly supercritical temperature, and to study the modifications brought about by a constant-density cooling down to liquid temperatures not far from the triple point. The multi-exponential analysis thus confirms its validity as a powerful method of analysis of correlation functions of disordered systems. In addition, this study has revealed the wealth of information contained in the VAF, due to its underlying relationship to all kinds of dynamical processes taking place in a system of particles undergoing various kinds of both collective and independent motions.

A great step forward with respect to the previous work consists in the identification of the nature of the various modes obtained through the fit of the exponential model. In fact, the modification of the fit model required at the various thermodynamic states has allowed us to understand the evolution of the VAF properties, providing essential suggestions for the definition of groups of modes to which clear and specific physical meaning could be attached. In this way we have identified the mode representation of three fundamental dynamical processes typical of a simple monatomic fluid, i.e.\ the long-time tail, the propagation of longitudinal sound waves, and the transverse dynamics with its clear transition to the excitation of propagating waves.

Our work confirms recent results on the existence of a crossover in the fluid state evidenced by the density evolution of the VAF time dependence. The multi-exponential analysis allowed us to bring to the foreground the role played here by the changing nature of transverse dynamics.

It is useful to assess the relative importance of the various processes as represented by their partial contributions to the total time-integrated VAF, reported in Table \ref{tab:integrals} as percent fractions, for all thermodynamic states. It is shown that the LTT contribution is the smallest one, being confined to the long time range where the VAF intensity decreases to very low values. The high frequency modes related to the sound propagation show a clear trend, growing in importance with both increasing density and decreasing temperature. Finally, the low frequency modes related to the transverse dynamics account for about half of the total VAF integral with no significant variation with the thermodynamic coordinates. This is interesting, because it evidences that the strength of the modes involved in the crossover does not show any abrupt change across the transition boundary.

Finally, we stress that in this work we demonstrated that the frequency spectrum of the VAF can be interpreted as a real density of states for the microscopic dynamics of a fluid.

\acknowledgments{We would like to thank Walter Penits for the maintenance of, and his kind help with, the HTCondor system  \cite{Condor}.}

\appendix
\section{}

We recall here from Ref.\ \cite{PRE2006} the main features of the VE model for the dynamic structure factor $S(Q,\omega)$, i.e. the time Fourier transform of the so-called intermediate scattering function $F(Q,t)$ which, in turn, is given by 
 
\begin{equation}
\label{Fdef}
F(Q,t)=\frac{1}{N} \sum_{\alpha,\beta=1}^{N} \langle e^{-i{\bf Q} \cdot {\bf R}_{\alpha} (0)} e^{i {\bf Q} \cdot {\bf R}_{\beta} (t)} \rangle
\end{equation}

\noindent and represents the time autocorrelation of the spatial Fourier components with wavevector ${\bf Q}$ of density fluctuations \cite{Balucani}. Here ${\bf R}_{\alpha}(t)$ is the position of the $\alpha$-th particle, and we exploit the isotropic behavior of a fluid to drop the dependence of $F(Q,t)$ on the direction of ${\bf Q}$.

The time evolution of $F(Q,t)$ can be described through the integro-differential equation

\begin{equation}
\label{Lang}
\ddot F(Q,t)+\int ^{t} _{0} \!dt' ~ K_2(Q,t-t')~\dot F(Q,t')~+\langle \omega_Q^2 \rangle~F(Q,t)=0,
\end{equation}

\noindent where the dots denote time derivatives, $\langle \omega_Q^2 \rangle=(k_{\rm B}TQ^2)/(mS(Q))$ is the normalized second frequency moment of $S(Q,\omega)$, $S(Q)$ is the static structure factor, and $K_2(Q,t)$ is the so-called second-order memory function in a hierarchy of equations describing the time dependence of the density autocorrelation \cite{Barocchi_2012}. Through the use of Laplace transforms to the complex frequency $s$ (here denoted by a tilde) and with initial conditions $F(Q,0)=S(Q)$ and $\dot F(Q,0)=0$, Eq. (\ref{Lang}) is solved to give

\begin{equation}
\label{contfrac}
\frac{\widetilde{F}(Q,s)}{S(Q)}=\left[ s+\frac{\langle \omega_{Q}^{2} \rangle}{s+\widetilde{K_2}(Q,s)}\right] ^{-1}.
\end{equation}

The viscoelastic model consists of taking $K_2(Q,t)$ in the form \cite{Balucani,PRE2006}

\begin{equation}
\label{memory}
\begin{split}
K_2(Q,t) = \Delta_{\rm L}^{2}(Q) {\rm exp}[-t/\tau(Q)]\\
+(\gamma(Q)-1)\langle \omega_{Q}^{2} \rangle {\rm exp}[-\Gamma_{\rm T}(Q) t],
\end{split}
\end{equation}

\noindent where $\Delta_{\rm L}^{2}(Q)=\omega_{\rm L}^{2}(Q)-\gamma(Q)\langle \omega_{Q}^{2} \rangle$, with $\omega_{\rm L}^{2}(Q)$ defined as the ratio of fourth to second spectral moment \cite{Balucani}. The quantities $\gamma(Q)$, $\Gamma_{\rm T}(Q)$ and $\tau(Q)$ are undetermined parameters to be fitted, which, however, in the hydrodynamic limit can be related to thermodynamic quantities, since for $Q\to 0$ one has $\gamma(Q)\to \gamma_0$, $\Gamma_{\rm T}(Q)/Q^2 \to \gamma_0D_{\rm T}$ and $\tau(Q)\to \tau_0$, where $\gamma_0$ is the ratio of the constant-pressure ($c_p$) to the constant-volume ($c_v$) specific heat, $D_{\rm T}$ is the thermal diffusivity, and $\tau_0$ is the $Q\to 0$ limit of $\nu Q^2/\Delta_{\rm L}^{2}(Q)$ with $\nu$ the kinematic longitudinal viscosity.

Substituting the Laplace transform of model (\ref{memory}) into Eq. (\ref{contfrac}) and following the steps described in Ref.\ \cite{PRE2006}, it can be shown that, at each $Q$,

\begin{equation}
\label{complexlor}
\frac{S(Q,\omega)}{S(Q)}=\frac{1}{\pi}{\rm Re}\frac{\widetilde{F}(Q,s=i\,\omega)}{S(Q)}=\frac{1}{\pi}{\rm Re}\sum_{j=1}^4 \frac{I_j}{i\,\omega-z_j},
\end{equation}

\noindent which is the spectrum of the normalized correlation function

\begin{equation}
\label{complexexp}
\frac{F(Q,t)}{F(Q,0)}=\sum_{j=1}^4 I_j \exp(z_j t)
\end{equation}

\noindent for $t\geq 0$. In both Eqs. (\ref{complexlor}) and (\ref{complexexp}) the index $j$ labels 4 terms described below. The latter equation shows that the VE model complies with the general theory of Sect.\ \ref{sect: multiexp}, although of course the $I_j$'s and $z_j$'s bear no relation to the analogous quantities obtained in the mode expansion of the VAF, since we are dealing here with the sum of exponentials relative to a different autocorrelation function. For ease of notation, we are omitting to indicate the explicit $Q$-dependence of $I_j$ and $z_j$ in Eqs. (\ref{complexlor})-(\ref{isf}).

Equation (\ref{complexexp}) describes two different dynamical situations depending on whether the sound modes are propagating or not. In the first case, two of the four terms of the sum are real and represent exponential decays but $F(Q,t)$ also contains a pair of complex conjugate terms providing a damped oscillatory part. The second case occurs in a rather narrow $Q$ range around $Q_{\rm p}$ where simple liquids show an arrest of sound propagation (called ``propagation gap'') which is reflected in the fact that the oscillating part of $F(Q,t)$ becomes overdamped, and all four modes of Eq. (\ref{complexexp}) are real.

Since we are here applying the VE model to the determination of the sound mode dispersion curve, the case of interest is the first one. Then, writing $z_j=-z_{\rm s}\pm i\,\omega_{\rm s}$ and $I_j=I_{\rm s}(1 \mp b_{\rm s})$ for the widths and amplitudes of complex conjugate mode pair, Eq. (\ref{complexlor}) leads to \cite{PRE2006}

\begin{equation}
\label{visco}
\begin{split}
S(Q,\omega) = \frac{S(Q)}{\pi} \Bigg[ I_1 \frac{|z_1|}{z_1^{2}+\omega^{2}} +I_2 \frac{|z_2|}{z_{2}^{2}+\omega^{2}}\\
+I_{\rm s} \frac{z_{\rm s}+b_{\rm s}(\omega+\omega_{\rm s})}{z_{\rm s}^{2}+ (\omega+\omega_{\rm s})^{2}}+I_{\rm s} \frac{z_{\rm s}-b_{\rm s}(\omega-\omega_{\rm s})}{z_{\rm s}^{2}+ (\omega-\omega_{\rm s})^{2}} \Bigg].
\end{split}
\end{equation}
 
\noindent  Here, subscripts 1 and 2 label the amplitudes and half-widths-at-half-maximum of two central Lorentzian lines, which represent the purely decaying modes and determine together the shape of the central peak. The subscript `s', meaning {\it sound}, refers to a pair of side lines centered at the excitation frequencies $\pm \omega_{\rm s}$ and having the shape of asymmetrically distorted Lorentzians which are the spectral representation of the two complex modes.

The spectrum (\ref{visco}) is the FT of the intermediate scattering function

\begin{equation}
\label{isf}
\frac {F(Q,t)}{F(Q,0)} = I_1~ e^{-|z_1| t} + I_2 ~ e^{-|z_2| t}+2 I_{\rm s}~ e^{-z_{\rm s} t} ~ \frac{{\rm cos}(\omega_{\rm s} t -\phi)}{{\rm cos}~\phi} 
 \end{equation}

\noindent for $t\geq 0$, which shows explicitly the presence of two exponentially-decaying terms plus an exponentially-modulated oscillation, where the phase angle $\phi$ is given by ${\rm tan}~\phi=b_{\rm s}$.

The VE model (\ref{memory}) complies with hydrodynamic results for $Q \to 0$, but exploits the presence of the relaxation time $\tau$ which is a characteristic feature of the concept of viscoelasticity. In the viscoelastic framework, the propagation speed of acoustic excitations is expected to undergo a transition from the low-$Q$ adiabatic sound velocity $c_{\rm s}$ to a higher value appropriately called the infinite-frequency speed $c_{\infty}$. The rationale behind such a prediction is the fact that when the excitation frequency grows with $Q$ up to values larger than the inverse of the relaxation time $\tau(Q)$, the relaxation mechanism quickly loses effectiveness giving the system a more rigid, i.e. ``elastic'', character. This phenomenon is one of the causes of the bending upwards of the dispersion curve, whose overall shape is however determined by a number of facts \cite{PRE2006}. In particular, the growth of the damping $z_{\rm s}$ with $Q$ tends instead to reduce the excitation frequency up to the point where the acoustic oscillations are brought into the overdamping condition causing the above mentioned propagation gap \cite{Sampoli_2008, Sampoli_2009}.


\begin{thebibliography}{}

\bibitem{Hansen} J. P. Hansen and I. R. McDonald, {\it Theory of Simple Liquids} (Academic Press, London, 1986).

\bibitem{Balucani} U. Balucani and M. Zoppi, {\it Dynamics of the Liquid State} (Clarendon, Oxford, 1994).

\bibitem{Boon} J. P. Boon and S. Yip, {\it Molecular Hydrodynamics} (Dover, New York, 1980).

\bibitem{Gaskell1} T. Gaskell and S. Miller, J. Phys. C: Solid State Phys. {\bf 11}, 3749 (1978).

\bibitem{Gaskell2} T. Gaskell and S. Miller, J. Phys. C: Solid State Phys. {\bf 11}, 4839 (1978).

\bibitem{Levesqueverlet} D. Levesque and L. Verlet, Phys. Rev. A {\bf 2}, 2514 (1970).

\bibitem{Alder_vaf} B. J. Alder and T. E. Wainwright, Phys. Rev. Lett. {\bf 18}, 988 (1967).

\bibitem{Alder_LTT} B. J. Alder and T. E. Wainwright, Phys. Rev. A {\bf 1}, 18 (1970).

\bibitem{Levesque} D. Levesque and W. T. Ashurst, Phys. Rev. Lett. {\bf 33}, 277 (1974).

\bibitem{McDonough} A. McDonough, S. P. Russo and I. K. Snook, Phys. Rev. E {\bf 63}, 026109 (2001).

\bibitem{Dib} R. F. A. Dib, F. Ould-Kaddour and D. Levesque, Phys. Rev. E {\bf 74}, 011202 (2006).

\bibitem{Meier1} K. Meier, A. Laesecke and S. Kabelac, J. Chem. Phys. {\bf 121}, 3671 (2004).

\bibitem{Meier2} K. Meier, A. Laesecke and S. Kabelac, J. Chem. Phys. {\bf 121}, 9526 (2004).

\bibitem{Williams} S. R. Williams, G. Bryant, I. K. Snook and W. van Megen, Phys. Rev. Lett. {\bf 96}, 087801 (2006).

\bibitem{Ryltsev} R. E. Ryltsev and N. M. Chtchelkatchev, J. Chem. Phys. {\bf 141}, 124509 (2014). 

\bibitem{Kawasaki} K. Kawasaki, Phys. Lett. {\bf 32A}, 379 (1970).

\bibitem{Ernst1970} M. H. Ernst, E. H. Hauge and J. M. J. van Leeuwen, Phys. Rev. Lett. {\bf 25}, 1254 (1970).

\bibitem{Ernst1971} M. H. Ernst, E. H. Hauge and J. M. J. van Leeuwen, Phys. Rev. A {\bf 4}, 2055 (1971).

\bibitem{Dorfman} J. R. Dorfman and E. G. D. Cohen, Phys. Rev. A {\bf 6}, 776 (1972).

\bibitem{paper1} S. Bellissima, M. Neumann, E. Guarini, U. Bafile and F. Barocchi, Phys. Rev. E {\bf 92}, 042166 (2015).

\bibitem{Isobe} M. Isobe, Phys. Rev. E {\bf 77}, 021201 (2008).

\bibitem{Barocchi_2012} F. Barocchi, U. Bafile and M. Sampoli, Phys. Rev. E {\bf 85}, 022102 (2012).

\bibitem{Barocchi_2013} F. Barocchi and U. Bafile, Phys. Rev. E {\bf 87}, 062133 (2013).

\bibitem{Barocchi_2014} F. Barocchi, E. Guarini and U. Bafile, Phys. Rev. E {\bf 90}, 032106 (2014).

\bibitem{Simeoni} G. G. Simeoni, T. Bryk, F. A. Gorelli, M. Krisch, G. Ruocco, M. Santoro and T. Scopigno, Nature Physics {\bf 6}, 503 (2010).

\bibitem{Gorelli} F. A. Gorelli, T. Bryk, M. Krisch, G. Ruocco, M. Santoro and T. Scopigno, Sci. Rep. {\bf 3}, 1203 (2013).

\bibitem{Bolmatov} D. Bolmatov, M. Zhernenkov, D. Zav'yalov, S. N. Tkachev, A. Cunsolo and Y. Q. Cai, Sci. Rep. {\bf 5}, 15850 (2015).

\bibitem{Cunsolo_2016} A. Cunsolo, Appl. Sci. {\bf 6}, 64 (2016).

\bibitem{Brazhkin_2012} V. V. Brazhkin, Yu. D. Fomin, A. G. Lyapin, V. N. Ryzhov and K. Trachenko, Phys. Rev. E {\bf 85}, 031203 (2012).

\bibitem{Brazhkin_2013} V. V. Brazhkin, Yu. D. Fomin, A. G. Lyapin, V. N. Ryzhov, E. N. Tsiok and K. Trachenko, Phys. Rev. Lett. {\bf 111}, 145901 (2013).

\bibitem{Allen} M. P. Allen and D. J. Tildesley, {\it Computer Simulation of Liquids} (Clarendon Press, Oxford, 1987).

\bibitem{Erpenbeck} J. J. Erpenbeck and W. W. Wood, Phys. Rev. A {\bf 26}, 1648 (1982).

\bibitem{Johnson} J. K. Johnson, J. A. Zollweg and K. E. Gubbins, Mol. Phys. {\bf 78}, 591 (1993).

\bibitem{Meier3} K. Meier, PhD thesis, University of the Federal Armed Forces Hamburg, 2002.

\bibitem{PRE2006} U. Bafile, E. Guarini and F. Barocchi, Phys. Rev. E {\bf 73}, 061203 (2006).

\bibitem{Boonpage} See Ref.\ \cite{Boon}, Ch. 3, page 123.

\bibitem{Glyde} H. R. Glyde, J. Low Temp. Phys. {\bf 59}, 561 (1985).

\bibitem{Rahman} A. Rahman, M. J. Mandell and J. P. McTague, J. Chem. Phys. {\bf 64}, 1564 (1976).

\bibitem{Garberoglio} G. Garberoglio and R. Vallauri, J. Chem. Phys. {\bf 115}, 395 (2001).

\bibitem{Cao} J. Cao and G. A. Voth, J. Chem. Phys. {\bf 103}, 4211 (1995).

\bibitem{Zanatta} M. Zanatta, F. Sacchetti, E. Guarini, A. Orecchini, A. Paciaroni, L. Sani and C. Petrillo, Phys. Rev. Lett. {\bf 114}, 187801 (2015).

\bibitem{Guarini}  E. Guarini, U. Bafile, F. Barocchi, A. De Francesco, E. Farhi, F. Formisano, A. Laloni, A. Orecchini, A. Polidori, M. Puglini and F. Sacchetti, Phys. Rev. B {\bf 88}, 104201 (2013).

\bibitem{Ashcroft} N. W. Ashcroft and N. D. Mermin, {\it Solid State Physics} (Saunders College, Philadelphia, 1976).

\bibitem{Sampoli} M. Sampoli, G. Ruocco and F. Sette, Phys. Rev. Lett. {\bf 79}, 1678 (1997).

\bibitem{Cunsolo_2012} A. Cunsolo, C. N. Kodituwakku, F. Bencivenga, M. Frontzek, B. M. Leu and A. H. Said, Phys. Rev. B {\bf 85}, 174305 (2012).

\bibitem{currents} For the definition of the current correlation functions we follow Ref.\ \cite{Balucani} and write $C_{\rm T}(Q,t)=1/(2N)\langle{\bf j}_{\rm T}^*({\bf Q},0) \cdot {\bf j}_{\rm T}({\bf Q},t) \rangle$, where ${\bf j}_{\rm T}({\bf Q},t)={\bf j}({\bf Q},t)-{\bf j}_{\rm L}({\bf Q},t)$, ${\bf j}_{\rm L}({\bf Q},t)=({\bf j}({\bf Q},t)\cdot {\bf Q}){\bf Q}/Q^2$ and finally ${\bf j}({\bf Q},t)=\sum_\alpha{\bf v}_\alpha(t)\exp[i{\bf Q}\cdot{\bf R}_\alpha(t)]$ where ${\bf R}_\alpha(t)$ and ${\bf v}_\alpha(t)$ are position and velocity of the $\alpha$-th particle. 

\bibitem{Fomin_2016} Yu. D. Fomin, V. N. Ryzhov, E. N. Tsiok, V. V. Brazhkin and K. Trachenko, J. Phys.: Condens. Matter {\bf 28}, 43LT01 (2016).

\bibitem{Frenkel} J. Frenkel, {\it Kinetic Theory of Liquids} (Oxford University Press, London, 1947).

\bibitem{Condor} \url{http://research.cs.wisc.edu/htcondor/}

\bibitem{Sampoli_2008} M. Sampoli, U. Bafile, F. Barocchi, E. Guarini and G. Venturi, J. Phys.: Condens. Matter {\bf 20}, 104206 (2008).

\bibitem{Sampoli_2009} M. Sampoli, U. Bafile, E. Guarini and F. Barocchi, Phys. Rev. B {\bf 79}, 214203 (2009).

\bibitem{Mastny} E. A. Mastny and J. J. de Pablo, J. Chem. Phys. {\bf 127}, 104504 (2007).

\end{thebibliography}
\end{document}